\documentclass[twocolumn,superscriptaddress]{revtex4}
\usepackage{dcolumn}
\usepackage{graphicx}
\usepackage{latexsym}
\usepackage{amsfonts}
\usepackage{amssymb}
\usepackage{dsfont}
\usepackage{ulem}

\usepackage{amssymb,amsmath,stmaryrd}
\usepackage{times}
\usepackage{float}
\usepackage{amsmath}
\usepackage{pst-node}
\usepackage{makeidx}
\usepackage{bbm}
\usepackage{cancel}
\usepackage{amsthm}
\usepackage{textgreek}

\makeatletter
\newcommand{\subalign}[1]{%
  \vcenter{%
    \Let@ \restore@math@cr \default@tag
    \baselineskip\fontdimen10 \scriptfont\tw@
    \advance\baselineskip\fontdimen12 \scriptfont\tw@
    \lineskip\thr@@\fontdimen8 \scriptfont\thr@@
    \lineskiplimit\lineskip
    \ialign{\hfil$\m@th\scriptstyle##$&$\m@th\scriptstyle{}##$\crcr
      #1\crcr
    }%
  }
}
\makeatother

\DeclareGraphicsExtensions{.pdf,.gif,.jpg}

\newcommand{\be}{\begin{equation}}
\newcommand{\ee}{\end{equation}}
\newcommand{\bea}{\begin{eqnarray}}
\newcommand{\eea}{\end{eqnarray}}

\def\a{\alpha}
\def\b{\beta}
\def\g{\gamma}
\def\G{\Gamma}
\def\d{\delta}
\def\D{\Delta}
\def\e{\epsilon}

\def\th{\theta}

\def\m{\mu}
\def\n{\nu}
\def\c{\xi}

\def\p{\pi}

\def\r{\rho}
\def\s{\sigma}
\def\S{\Sigma}
\def\t{\tau}

\def\w{\omega}
\def\W{\Omega}

\def\blk{{\mathbf k}}

\def\blp{{\mathbf p}}
\def\blq{{\mathbf q}}

\def\callA{\mbox{$\mathcal{A}$}}

\def\callN{\mbox{$\mathcal{N}$}}
\def\callO{\mbox{$\mathcal{O}$}}

\def\callT{\mbox{$\mathcal{T}$}}

\def\bra{\langle}
\def\ket{\rangle}

\def\Re{{\rm Re}}

\def\1op{\hat{\mathbbm{1}}}

\def\AA{\mathring{\mathrm{A}}}

\begin{document}

\title{Ultrafast creation and disruption of excitonic condensate in
transition metal dichalcogenides revealed by time-resolved ARPES}

\title{Transient exciton condensation and its ultrafast 
disruption in  transition metal dichalcogenides revealed by time-resolved ARPES}

\title{Ultrafast melting of the nonequilibrium excitonic insulator 
phase in bulk WeSe$_{2}$}

\title{Ultrafast creation and melting of nonequilibrium excitonic 
condensates  in bulk WSe$_{2}$}

\author{E. Perfetto}
\affiliation{Dipartimento di Fisica, Universit\`{a} di Roma Tor Vergata,
Via della Ricerca Scientifica 1, 00133 Rome, Italy}
\affiliation{INFN, Laboratori Nazionali di Frascati, Via E. Fermi 40, 00044 Frascati, 
Italy}

\author{G. Stefanucci}
\affiliation{Dipartimento di Fisica, Universit\`{a} di Roma Tor Vergata,
Via della Ricerca Scientifica 1, 00133 Rome, Italy}
\affiliation{INFN, Laboratori Nazionali di Frascati, Via E. Fermi 40, 00044 Frascati, 
Italy}

\begin{abstract}

We study the screened dynamics of the nonequilibrium excitonic consensate forming in a bulk 
WSe$_{2}$ when illuminated by coherent light resonant 
with the lowest-energy exciton. Intervalley scattering causes 
electron migration from the optically populated K valley to the 
conduction band minimum at $\S$. Due to the  electron-hole unbalance 
at the K point a plasma of quasi-free holes develops, which
efficiently screens the interaction of the remaining excitons.
We show that this plasma screening causes an ultrafast melting of the 
nonequilibrium consensate and that during melting coherent excitons 
and quasi-free electron-hole pairs coexist. The time-resolved 
spectral function does exhibit a conduction  
and excitonic sidebands of opposite convexity and   
relative spectral weight that changes 
in time. Both the dependence of 
the time-dependent conduction density on the laser intensity and the 
time-resolved spectral function agree with recent experiments.

\end{abstract}

\maketitle

\section{Introduction}

Excitons in solids are electron-hole (e-h) bound pairs 
which behave as composite bosons  in 
the dilute limit~\cite{keldysh1972problems,Keldysh-Kozlov_JETP1968,2000bceb,combescot2016excitons}.
More than three decades ago it was suggested that a 
nonequilibrium (NEQ) 
exciton superfluid may form in a semiconductor after pumping with coherent light of 
frequency smaller than the gap but larger than or at most equal to 
the exciton 
energy~\cite{Schmitt-Rink_PhysRevB.37.941,kuklinski1990,Glutsch_PhysRevB.45.5857,littlewood1996}. 
The pump would then drive the system from a 
non-generate ground-state (insulating phase) 
to a symmetry-broken excited state known as the NEQ excitonic 
insulator 
(EI)~\cite{Ostreich_1993,Hannewald-Bechstedt_2000,GLUTSCH1992,PSMS.2019}. 
It has been recently shown that the NEQ-EI state emerges 
also from the spontaneous symmetry breaking of a macroscopically 
degenerate excited 
manifold~\cite{PSMS.2019,SzymaPRL2006,Hanai2016,HanaiPRB2017,TriolaPRB2017,Hanai2018,PertsovaPRB2018,Becker_PhysRevB.99.035304,Yamaguchi_PhysRevLett.111.026404,pertsova2020}.

In time-resolved (tr) ARPES the NEQ-EI phase generates 
a replica of the valence band at the exciton energy, i.e. {\it below} 
the conduction band minimum 
(CBM)~\cite{Schmitt-Rink_PhysRevB.37.941,PSMS.2019,PhysRevB.100.205401,PBS.2020}.
The experimental observation of this excitonic sideband is usually
challenging~\cite{madeo2020,lee2020} since excitons may quickly loose 
coherence~\cite{Koch2006} due to 
electron-phonon scattering~\cite{madrid2009,nie2014,Selig2016,sangalli2018ab} or break due
to excited-state screening 
mechanisms~\cite{Chernikov2015,Chernikov20152,Cunningham2017,Yao2017,WangArdelean2019,Dendzik2020prl}.
As a result the ARPES signal changes rapidly with increasing pump-probe 
delay~\cite{PSMS.2016,Steinhoff2017,RustagiKemper2018,PhysRevB.100.205401,PBS.2020} and the measured 
spectra become difficult to interpret.
The development of a microscopic theory which 
takes into account decoherence, screening and
other material-specific properties is necessary in order to 
understand the ultrafast dynamics and interpret the experimental 
results.

\begin{figure}[tbp]
    \includegraphics[width=0.45\textwidth]{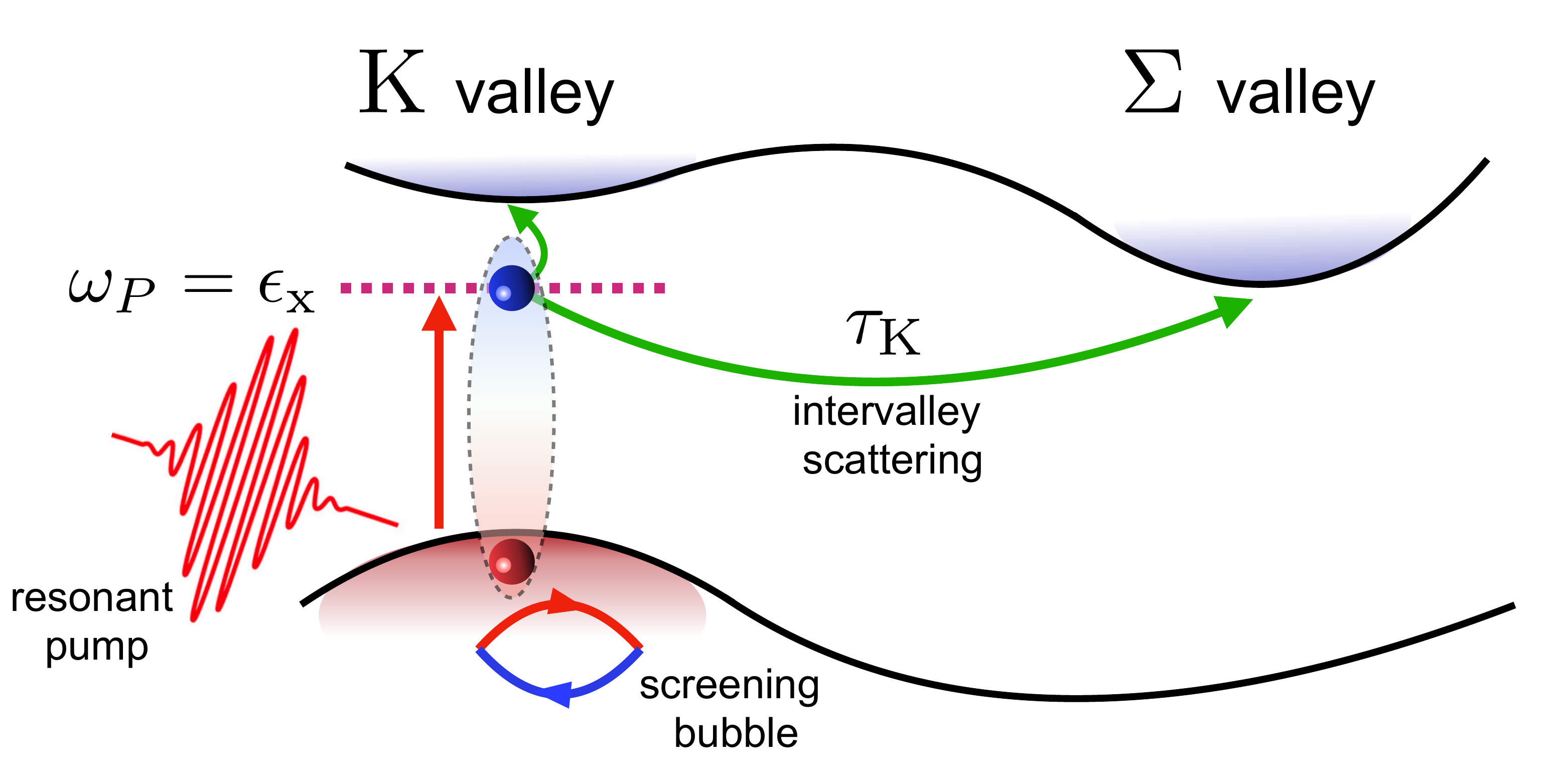}
 \caption{The NEQ-EI state is 
 created in the K valley, where bright excitons are initially 
 photo-excited. Subsequent  $\mathrm{K}\to \S$ intervalley scattering 
 of excited carriers 
 breaks charge neutrality in the K valley, and free K-valence holes 
 screen the Coulomb repulsion. The reduction of e-h attraction 
 causes exciton breaking, and free electrons occupy the available conduction 
 states in the K valley.}
 \label{model}
 \end{figure}

In this work we put forward a microscopic theory for bulk  WSe$_{2}$, an 
indirect gap semiconductor~\cite{Jiang2012} with an optically bright exciton of energy slightly 
above the gap~\cite{beal1976,finteis1997,arora2015,riley2015,kim2016}. A distinct physical picture emerges from our 
calculations, see Fig.~\ref{model}. The pump-induced photoexcitation initially generates an 
exciton superfluid around the K point~\cite{frindt1963,WangChernikov2018}. Due to intervalley 
scattering~\cite{Bertoni_PhysRevLett.117.277201}, 
however, electrons migrate from the 
K-valley to the CBM at the $\S$-valley~\cite{Jiang2012}, and excitons begin to 
dissociate. Consequently, a gas of free holes, i.e., holes not bounded 
to conduction electrons, forms in the valence band at the K-point.
This hole-plasma efficiently screens~\cite{Steinhoff2014,Liang2015,Meckbach2018} the electron-hole attraction
and eventually causes an ultrafast melting of the remaining 
excitons~\cite{Dendzik2020prl}. During the melting process the quasi-free 
e-h pairs coexist with 
excitons in the NEQ-EI phase.

Through real-time simulations and self-consistent excited-state 
calculations we show the formation of the  
NEQ-EI phase during resonant pumping as well as its fingerprints in 
tr-ARPES spectra. If screening is 
neglected then the excitonic sideband is simply 
attenuated by the K$\to \S$ intervalley scattering
as the pump-probe delay increases. If, instead,  both 
screening and intervalley scattering  are taken into account then 
the instantaneous formation of the excitonic sideband is followed by 
the development of a  
quasiparticle sideband, signaling the coexistence of 
quasi-free e-h pairs and excitons. 
The spectral weight is then
rapidly transferred from the excitonic sideband to the quasiparticle 
sideband until the extinction of the former. 
The calculated tr-ARPES spectra of bulk $\mathrm{WSe}_{2}$ agrees well 
with recent experiments on the same system~\cite{Puppin-PhD}
if the K$\to \S$ scattering rate is estimated to be $\approx 
15$~fs~\cite{Bertoni_PhysRevLett.117.277201,Dendzik2020prl,Puppin-PhD}.
Our simulations indicate that in this case
the NEQ-EI phase melt in few tens of femtoseconds.
Although the interplay of intervalley scattering and enhanced 
screening is investigated in bulk $\mathrm{WSe}_{2}$ the  
highlighted mechanism is general and it is likely to occur in other indirect gap 
semiconductors.

The paper is organized as follows. In the next Section we present 
our microscopic modelling of bulk $\mathrm{WSe}_{2}$. 
We illustrate
the equation of motion for the one-particle density matrix 
that accounts for coherent exciton formation in the K 
valley upon optical excitation, for the subsequent K$\to\S$ intervalley 
scattering of excited carriers, and for the plasma-induced screening.
In Section~\ref{sec3} we discuss the Floquet solution of the  equation 
of motion, useful for characterizing the transient NEQ-EI state at 
the K point 
forming after resonant pumping.
Real-time simulations of WSe$_{2}$ driven by a VIS pulse of duration 
$\simeq 10^{2}$~fs are introduced  
in Section~\ref{sec4}. In Section~\ref{sec5} we present 
results in the absence of plasma screening. We show that in this case 
the time-dependent solution follows adiabatically the Floquet solution
and that exciton dissociation induced 
by intervalley scattering does not destroy the NEQ-EI phase. 
In Section~\ref{sec6} both intervalley scattering and screening 
effects are taken into account. Through tr-ARPES spectra we show that the initial NEQ-EI phase 
is slowly contaminated by quasi-free e-h pairs until an incoherent plasma of 
electrons and holes is formed. 
A summary of the main findings are drawn in Section~\ref{sec7}.

\section{Microscopic Theory}
\label{sec2}

Bulk $\mathrm{WSe}_{2}$  is an indirect-gap semiconductor~\cite{Jiang2012} with the 
property that 
optical excitations of frequency close or below the gap
create e-h pairs around the  K and K' 
points~\cite{frindt1963,WangChernikov2018}, where the direct  gap is located~\cite{Jiang2012}.
The two-dimensional (2D) character 
of the valence and conduction band close to the K points~\cite{Bertoni_PhysRevLett.117.277201} allows for 
modelling the e-h dynamics using a 2D two-band model.
In the following we neglect the spin-orbit coupling and consider 
degenerate and decoupled K and K' 
valleys. The inclusion of the spin-orbit coupling does not change the 
main conclusions.
Let $\e_{v \blk}$ and $\e_{c \blk}$ be the valence and conduction 
dispersion and $\e_{g} \equiv
\mathrm{min}[\e_{c \blk}]-\mathrm{max}[\e_{v\blk}]=1.8$~eV~\cite{Jiang2012,beal1976,finteis1997,arora2015,riley2015,kim2016} define the direct gap. 
Placing the K point at $\blk=0$ we
use quadratic dispersions $\e_{v \blk}=-\frac{k^{2}}{2m}-\frac{\e_{g}}{2}$ and 
$\e_{c \blk}=\frac{k^{2}}{2m}+\frac{\e_{g}}{2}$, with 
$k=|\blk|$. 
We parametrize the electron-hole attraction according to Ref.~\cite{Steinhoff2017}
\be
U_{\blq}=\frac{v_{\blq}}{\e_{\blq}} ,
\ee
where
\begin{align}
 v_{\blq}&=\frac{2\p}{q(1+\g q + \d q^{2})}
\end{align}
is the bare interaction and 
\begin{align}
 \e_{\blq}&=\e^{\infty}_{\blq}\frac{1-\b_{1 
 \blq}\b_{2\blq}e^{-2hq}}{1+(\b_{1 \blq}+\b_{2 \blq})e^{-hq}+\b_{1 
 \blq}\b_{2\blq}e^{-2hq}}
 \label{espilonq}
\end{align}
accounts for the screening. 
In Eq.~(\ref{espilonq}) 
\begin{align}
 \e^{\infty}_{\blq} &= g+\frac{a+q^{2}}{\frac{a\sin(qc)}{qbc}+q^{2}}
\end{align}
is the dielectric function of the neglected bands whereas 
\begin{align}
 \b_{i \blq}  &=\frac{\e^{\infty}_{\blq}-\e_{\mathrm{sub,i}}}{\e^{\infty}_{\blq}+\e_{\mathrm{sub,i}}}
\end{align}
takes into account the dielectric constant $\e_{\mathrm{sub,i}}$ of a 
possible substrate ($i=1$) or superstrate ($i=2$). 
Realistic parameters to describe a single layer of $\mathrm{WSe}_{2}$ 
close to the K point 
are~\cite{Steinhoff2017} $m/m_{e}=0.34$ (where $m_{e}$ is the free electron mass),
$a=2.4~\AA^{-2}$, $b=21$, $c=5.7~\AA$, 
$h=2.5~\AA$, $g=5.3$, 
$\g=2.3~\AA$, $\d=0.174~\AA^{2}$.
Top and bottom layers play the role of a superstrate and substrate 
respectively~\cite{PhysRevB.92.245123};  therefore 
$\e_{\mathrm{sub,1}}=\e_{\mathrm{sub,2}}\equiv \e$.
The value  $\e=6$ is chosen to reproduce 
the lowest (bright) A-exciton of energy $\e_{\rm x} = 1.7$~eV  
(binding energy $\e_{b}=\e_{g}-\e_{\rm x}=0.1~\mathrm{eV}$), in  
agreement with the literature~\cite{beal1976}.

We are interested in the electronic properties of $\mathrm{WSe}_{2}$
under weak resonant pumping -- hence with a photon frequency  
equal to the lowest bright exciton energy.
As we shall see the injected exciton fluid inherits the 
coherence of the laser pulse and a NEQ-EI phase is transiently 
generated. Pump-probe 
experiments~\cite{wallauer2016,Bertoni_PhysRevLett.117.277201,waldecker2017,madeo2020} 
have provided evidence that 
excited carriers experience a fast 
intervalley scattering due to 
electron-electron~\cite{HaugKochbook,Steinhoff2016,schmidt2016} and electron-phonon 
interactions~\cite{Selig2016,MolinaSanchez2017}. The intervalley 
scattering transfers the pumped electrons from the K valley to the 
CBM at the $\S$ valley on a time-scale $\t_{\mathrm{K}} \approx 15$~fs~\cite{Puppin-PhD}.
After the K$\to\S$ scattering electrons escape from the 
considered layer since the valence band at the
$\S$ point has a three-dimensional character~\cite{Bertoni_PhysRevLett.117.277201}.
The lickage of conduction electron from the K point 
is taken into account by adding a drain term to the equation of motion 
for the density matrix, see below.

Intervalley scattering has also a pivotal role in renormalizing the effective
 e-h attraction.
Indeed the total conduction density $n_{\rm e}(t)$ (i.e. the sum of K and K' 
valleys contributions) becomes smaller than the corresponding valence 
hole-density 
$n_{\rm h}(t)$. This gives rise to 
a finite density $n_{\mathrm{pl}}=(n_{\mathrm{h}}-n_{\rm e})/2 $ of free 
holes in each K valley. Under the weak pumping assumption
the screening due to excitons is negligible
and can be discarded~\cite{PhysRevB.102.085203}.
Thus the screened e-h interaction $W_{\blq}$ is given by
\be
W_{\blq}=\frac{U_{\blq}}{1-2U_{\blq}
\chi_{ \blq}^{\mathrm{pl}}},
\label{hpsex}
\ee
where $\chi_{ \blq}^{\mathrm{pl}}$ is 2D Lindhard function~\cite{GiulianiVignale-book} 
\be
\chi_{ \blq}^{\mathrm{pl}}=\frac{m}{\pi}\left[\theta(q-\sqrt{8\pi 
n_{\mathrm{pl}}})\sqrt{1-\frac{8\pi 
n_{\mathrm{pl}}}{q^{2}}}-1\right].
\label{xpsex}
\ee

With this premise, the equation of motion for the $2\times 2$
one-particle density matrix in the 
Hartree plus Screened Exchange (HSEX) approximation reads
\be
-i\frac{d}{dt}\r_{\blk}(t)+[h_{\mathrm{HSEX}, \blk}(t),\r_{\blk}(t)] 
-\frac{i}{2} 
\{\G, \r_{\blk}(t) \}=0,
\label{eomtdhpsex}
\ee
where the $2\times 2$ matrix of the drain term 
\be
\G=\left(
\begin{array}{cc}
  0 &  0 \\ 
    0 & \t^{-1}_{\mathrm{K}}   \\ 
\end{array}
\right)
\label{gammamat}
\ee
accounts for intervalley scattering and it 
ensures an exponential depletion of the conduction density on the $\t_{\mathrm{K}}$
time-scale.
The time-dependent HSEX hamiltonian in the presence of an electric 
field $E(t)$ coupled to the valence-conduction dipole moments $d_{\blk}$
reads
\be
h_{\mathrm{HSEX}, \blk}(t) = \left(
\begin{array}{cc}
  \e_{v \blk}+   V^{vv}_{\blk}(t)  &  
  V^{vc}_{ \blk}(t) +E(t) d_{\blk} \\ 
    V^{cv}_{ \blk}(t) +  E(t) d_{\blk} & \e_{c \blk}+ 
    V^{cc}_{ \blk}(t)   \\ 
\end{array}
\right)
\ee
with HSEX potential 
\begin{subequations}
\begin{align}
V^{vv}_{ \blk}(t) & =  -\frac{1}{\callN} \sum_{\blq} 
W_{\blk-\blq}(t)
 [\r^{vv}_{\blq}(t)-1]  \\
 V^{cc}_{ \blk}(t) & =  -\frac{1}{\callN} \sum_{\blq} 
W_{\blk-\blq}(t)
 \r^{cc}_{\blq}(t)  \\
V^{cv}_{ \blk}(t) &= V^{vc *}_{ \blk}(t)=
-\frac{1}{\callN} \sum_{\blq} W_{\blk-\blq}(t)
 \r^{vc}_{ \blq}(t).
 \label{VHSEXcv}
\end{align}
\end{subequations}
We emphasize that the HSEX potential depends on the density matrix explicitly as 
well as implicitly through 
the dependence of $W$ on $n_{\mathrm{pl}}=(n_{\rm h}-n_{\rm e})/2$. 
The density of conduction electron and valence holes  is indeed given 
by  $n_{\rm e}(t)=\frac{4}{\mathcal{NA}}\sum_{\blk}\r^{cc}_{\blk}(t)$ and 
$n_{\rm h}(t)=\frac{4}{\mathcal{NA}}\sum_{\blk}(1-\r^{vv}_{\blk}(t))$, 
where $\mathcal{N}$ is the number of $\blk$-points and $\mathcal{A}=9.53 \times 10^{-16}$~cm$^{2}$ 
is the area of the unit 
cell of a $\mathrm{WSe}_{2}$ layer.
Here the factor 4 in front of $n_{\rm e}$ and $n_{\rm h}$ accounts for
the spin and valley degeneracy.

Below we solve numerically Eq.~(\ref{eomtdhpsex}) using 
the ground-state density matrix 
$\r_{g,\blk}=\left(\begin{array}{cc}1&0\\0&0\end{array}\right)$ as 
initial condition. However, in order to 
shed light on the nature of the time-dependent results
we preliminary discuss the Floquet solution of Eq.~(\ref{eomtdhpsex}) at zero 
electric field.

\section{NEQ-EI phase from the Floquet solution}
\label{sec3}

The Bose-Einstein condensation of excitons in equilibrium matter is 
predicted in narrow-gap
semiconductors or semimetals where the exciton binding  energy  is 
larger  than  the  
bandgap~\cite{HalperinRice1968,Blatt1962,KeldyshKopaev1965,Kozlov-Maksimov_JETP1965,JeromeRiceKohn1967,Keldysh-Kozlov_JETP1968,Comte-Nozieres_1982}.
In this case the system is unstable toward the spontaneous formation of excitons, leading to a
symmetry-broken state called excitonic insulator (EI).
The mean-field theory of EIs has a close analogy to the BCS theory
and the resulting exciton superfluid is characterized by a
nonvanishing order parameter $\rho^{cv}_{\blk}$.
In a NEQ-EI the symmetry breaking is induced by the pump field~\cite{Schmitt-Rink_PhysRevB.37.941}. 
However, we have recently 
shown that the  NEQ-EI phase can also be obtained by solving a BCS-like 
problem with different chemical potentials $\m_{v}$ and $\m_{c}$ 
for valence and conduction 
electrons~\cite{PSMS.2019}. Below we briefly revisit this excited-state 
self-consistent approach and provide a characterization of the 
solution. As we are interested in describing a stable NEQ-EI we 
discard the intervalley scattering for the time being. 

In the HSEX approximation the lowest-energy density matrix 
corresponding to a quantum state with a finite 
density of conduction electrons and 
valence holes can be written as~\cite{PSMS.2019,PhysRevB.102.085203}
\begin{align}
    \r^{\a\b}_{\blk}=\sum_{\c}f(e^{\c}_{\blk})\varphi^{\c}_{\a 
    \blk}\varphi^{\c\ast}_{\b \blk}.
    \label{neqeidm}
\end{align}
In Eq.~(\ref{neqeidm})  $f(x)=1/(e^{x/T}+1)$ is the Fermi function 
at  temperature $T$ and 
the two-dimensional vector  $\varphi^{\c}_{\blk}$  solves the secular 
problem
\be 
\big(h_{\rm 
HSEX,\blk}-\mu\big)\varphi^{\c}_{\blk}=e^{\c}_{\blk}\varphi^{\c}_{\blk},
\label{hhsex}
\ee
where $\c=\pm$ labels the two eigenvectors and 
\begin{align}
    \mu=\left(\!\!
\begin{array}{cc}
  \m_{v}  &  
 0 \\ 
   0&\m_{c} \\ 
\end{array}\!\!
\right)\!\!
\label{mu}
\end{align}
is the matrix of chemical potentials ensuring that both $n_{\rm e}$ 
and $n_{\rm h}$ are nonvanishing. 
The secular problem must be solved self-consistently since the 
HSEX potential is a functional of $\r$. The symmetry of the problem ensures that 
$e^{-}_{\blk}<0<e^{+}_{\blk}$, hence charge neutral solutions $n_{\rm 
e}=n_{\rm h}$ are 
obtained for $\m_{v}=-\m_{c}$. In the reminder of this work we work 
at zero temperature and therefore
$\r^{\a\b}_{\blk}=\varphi^{-}_{\a 
\blk}\varphi^{-\ast}_{\b \blk}$.
The ground state solution $\r=\r_{g}$ is recovered for 
$\m_{v}=\m_{c}=0$, as it should.

\begin{figure}[tbp]
\includegraphics[width=0.4\textwidth]{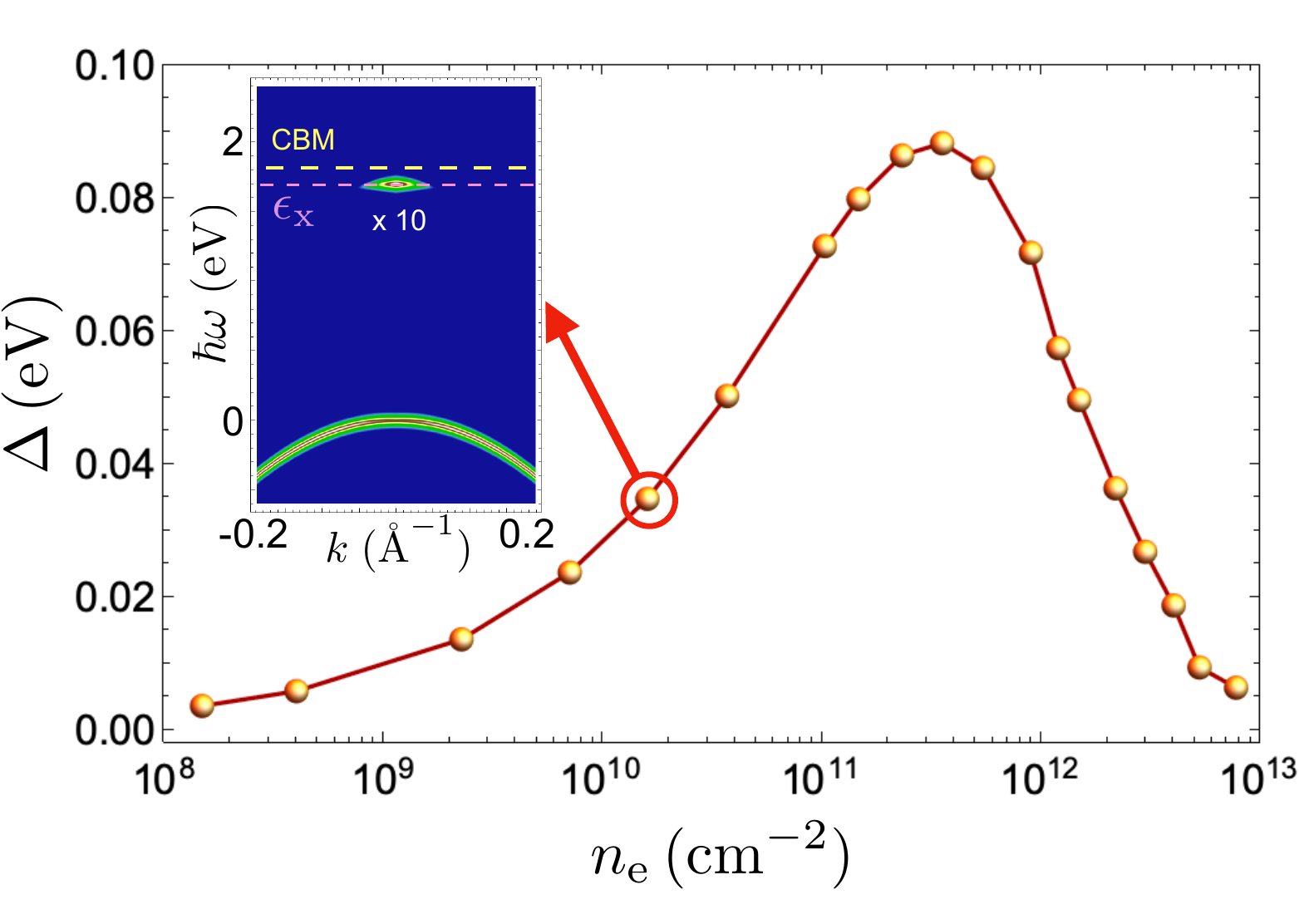}
\caption{Order parameter $\D$ versus
electronic density $n_{\rm e}$.
The inset shows the spectral function in Eq.~(\ref{spectral}) for 
$n_{\mathrm{e}}=1.6 \times 10^{10}\mathrm{cm}^{-2}$.
The excitonic sideband has been magnified by 
a factor 10 for a better visualization. Energies are measured with respect to the 
maximum of the valence band (VBM). Horizontal dashed lines at  
$\w = \e_{g}$ and $\w = \e_{\mathrm{\rm x}}$  are also drawn.  }
 \label{figphase}
 \end{figure} 

A symmetry broken 
NEQ-EI solution exists if the difference 
between the chemical potentials is larger than the lowest exciton energy 
$\e_{\mathrm{\rm x}}$, i.e.,  $\d \m \equiv 
\m_{c}-\m_{v}>\e_{\mathrm{\rm x}}$~\cite{PSMS.2019}. The NEQ-EI solution is characterized by a finite 
$\r_{\blk}^{cv}$ and an infinite degeneracy since if 
$\varphi^{\c}_{\blk}=\left(\begin{array}{c}\varphi^{\c}_{v\blk}\\\varphi^{\c}_{c\blk}\end{array}\right)$
is a solution then $\varphi^{\c}_{\blk}(\th)=
\left(\begin{array}{c}\varphi^{\c}_{v\blk}\\e^{i\th}\varphi^{\c}_{c\blk}\end{array}\right)$
is a solution too.
To quantify the magnitude of the 
symmetry breaking we choose the order parameter to be 
the HSEX potential in Eq.~(\ref{VHSEXcv})  calculated at $\blk=0$ and with real vectors 
$\varphi^{\c}_{\blk}$
\be
\D \equiv V^{cv}_{\blk=0}=- \frac{1}{\callN} \sum_{\blq} 
   W_{\blq}
  \varphi^{-}_{c \blq} \varphi^{-}_{v \blq} .
\label{orpar}
\ee
In Fig.~\ref{figphase} we show the values of $\D$ for the two-band 
model of $\mathrm{WSe}_{2}$. The order parameter is nonvanishing 
up to high densities
$n_{\rm e} \sim 10^{13}\mathrm{cm}^{-2}$, and it reaches its maximum 
value for $n_{\rm e} \sim 5 \times 10^{11}\mathrm{cm}^{-2}$.
We observe that in this calculation $n_{\rm pl}=0$ since no 
intervalley scattering is included. Hence $W_{\blq}=U_{\blq}$ 
coincides with the bare interaction,
in agreement with recent findings on excitonic 
screening, see Ref.~\cite{PhysRevB.102.085203} and Appendix A.

Useful insight on the self-consistent solution comes from the 
low-density limit, i.e., $ \d \m \gtrsim \e_{\rm x}$. It can be 
shown~\cite{PSMS.2019} that in this case 
\be
e^{-}_{\blk}\approx \e_{v \blk}+\d \m /2 ,
\label{linearre}
\ee
with 
\be
\left(\begin{array}{c}
\varphi^{-}_{v \blk}  \\ \varphi^{-}_{c \blk}
\end{array}\right)\simeq 
\left(\begin{array}{c}
1 \\ \sqrt{(N_{\rm e}/2)} \,Y_{\blk}
\end{array}\right)
\label{linearr}.
\ee
In Eq.~(\ref{linearr}) the quantity $N_{\rm e}=\frac{1}{2}\callN\callA\, n_{\rm e}$
is the number of conduction electrons 
per valley whereas $Y_{\blk}$ is the normalized, i.e., 
$\sum_{\blk}|Y_{\blk}|^{2}=1$,  lowest-energy solution of the Bethe-Salpeter equation
\be
(\e_{c \blk}-\e_{v \blk}-\e_{\rm x})Y_{\blk}=\frac{1}{\mathcal{N}}
\sum_{\blq} U_{\blk -\blq}Y_{\blq}.
\ee
Defining the creation operator for the (zero-momentum) 
exciton according to 
$\hat{b}^{\dag}=\sum_{\blk\s} Y_{\blk} \hat{c}^{\dag}_{\blk 
\s}\hat{v}_{\blk \s}$ we show in Appendix B  that 
total number of excitons $N_{\mathrm{ex}}\equiv \bra 
\hat{b}^{\dag} \hat{b} \ket$ 
is the same as $N_{\rm e}$, i.e.,
\be
N_{\mathrm{ex}} = N_{\rm e}.
\label{bec}
\ee
Thus in the dilute limit 
 {\it all} e-h pairs participate to the creation of 
excitons and these excitons condense in the lowest energy state, 
consistently with the BEC picture. 

The most remarkable feature of the NEQ-EI phase is that the 
corresponding quantum state is 
not a stationary state. In fact, using the NEQ-EI density matrix as 
initial condition one finds that the equation of 
motion~(\ref{eomtdhpsex}) (with $\G=0$) at zero external field, i.e., 
$E(t)=0$, is satisfied by 
\be
\r_{\blk}(t)=\left(\begin{array}{cc}
\r_{\blk}^{vv} & \r^{vc}_{\blk}e^{-i\d \m t} \\
\r^{cv}_{\blk}e^{i\d \m t} & \r_{\blk}^{cc}
\end{array}\right),
\ee
implying that  the order parameter  evolves in time according 
to~\cite{Ostreich_1993,SzymaPRL2006,PSMS.2019}
\be
 \D(t)=\D e^{-i\d \m t}.
 \label{persosc}
\ee
These self-sustained oscillations generate a Floquet-like 
regime~\cite{PhysRevLett.125.106401}
in the absence of external driving which is 
expected to survive over a timescale dictated by the exciton 
lifetime. Striking  features of the NEQ-EI phase have been 
predicted in relation to time-resolved (tr) ARPES 
experiments~\cite{PBS.2020}. 
For long enough probes the excitonic condensate generates a replica of the 
valence band at the exciton energy. Reducing the probe duration below the 
condensate period $T_{\rm NEQ-EI}\equiv 2\p/\d\m$ the replica fades away 
and the ARPES signal becomes 
periodic with period $T_{\rm NEQ-EI}$.
The observation of the latter effect is experimentally challenging. 
However, distinguishing the excitonic replica is within reach of 
modern ARPES techniques provided that the exciton life-time is longer 
than the inverse of the exciton energy. The ARPES signal is 
in this case proportional to the spectral function which is in turn 
given by~\cite{PSMS.2019}
\be
A_{\blk}(\w)=|\varphi^{-}_{v \blk}|^{2}\d(\w - e^{-}_{\blk} +
\frac{\d \m}{2})+|\varphi^{-}_{c \blk}|^{2}\d(\w - e^{-}_{\blk} - \frac{\d \m}{2}).
\label{spectral}
\ee
In the dilute limit
$e^{-}_{\blk}\approx \e_{v \blk}+\d \m /2$, see Eq.~(\ref{linearre}), and therefore a replica of the valence
band (shifted upward by $\e_{\rm x}$) appears 
inside the gap, see inset of Fig.~\ref{figphase}. 
Interestingly, the spectral weight of the
excitonic sideband is
proportional to the $\blk$-resolved conduction density  $|\varphi^{-}_{c 
\blk}|^{2}$ which, according to Eq.~(\ref{linearr}),
is proportional to the square of the excitonic wavefunction. 

In the following we show that the NEQ-EI phase 
discussed in this Section 
can be generated in real-time by driving the system with laser 
pulses of proper subgap frequency. The
inclusion of intervalley scattering  
and screening, however,  affect the stability of the photoinduced
exciton superfluid with inevitable and interesting repercussions on 
the  time-dependence of the spectral function.

\section{Real-time simulations}
\label{sec4}

We investigate the non-equilibrium  electronic properties of $\mathrm{WSe}_{2}$
under  {\it weak pumping resonant with the lowest bright exciton energy}.
As resonant photoexcitation provides an efficient injection of 
excitons, we expect that the NEQ-EI phase can be
reached during the time evolution.
Indeed, the external radiation transfers 
coherence to the e-h liquid which can then condense in an exciton 
superfluid.

In our simulations the system is photoexcited by a VIS 
pulse of finite duration $T_{P}$, maximum intensity $E_{P}$ and 
centered around the frequency $\w_{P}$:
\be
E(t)=\th(1-|1-2t/T_{P}|)
E_{P}\sin^{2}(\frac{\p t}{T_{P}})\sin(\w_{P}t),
\label{pulse}
\ee
To display our numerical results
it is convenient to set the origin of times at  $T_{P}/2$ (pump peak).
This choice is inspired by the experimental convention 
to set the origin of delays $\t$ when the temporal distance between
the pump and probe peaks vanishes.
Notice that this convention allows for a 
a direct comparison with  esperimental 
data since the calculated conduction density $n_{\rm e}(\t)$ at time $\t$ 
is proportional to the  
tr-ARPES spectral weight relative to the conduction band observed at 
delay $\t$. 
We  assume momentum-independent dipole moments $d_{\blk}=d$ 
and define the Rabi frequency, which determines the strength of the 
light-matter coupling, as $\Omega_{P}=E_{P}d$.
The simulations have been performed with the CHEERS code~\cite{PS-cheers} using a weak 
pump, $\Omega_{P}=13$~meV, of duration 
$T_{P}=100$~fs (FWHM = 50~fs) and resonant frequency 
$ \w_{P}=\e_{\rm x}=1.7$~eV.
According to our convention the initial time is then 
$t=-50$~fs.

We calculate the density of conduction electrons $n_{\rm e}(t)$ 
and valence holes $n_{\rm h}(t)$ as well as 
the time-dependent order parameter  
$\D(t)=-\frac{1}{\mathcal{N}}\sum_{\blq}W_{\blq}\r^{vc}_{\blq}(t)$.
We also calculate the time-dependent spectral function 
$A_{\blk}(\t,\w)$ at time $\t$.
As discussed in Ref.~\cite{PBS.2020} this quantity
is proportional to the tr-ARPES spectrum measured  
by a EUV probe at delay $\t$. In our simulations we have used a 
probing temporal window of 70~fs, see also Appendix C.

\begin{figure}[tbp]
\includegraphics[width=0.45\textwidth]{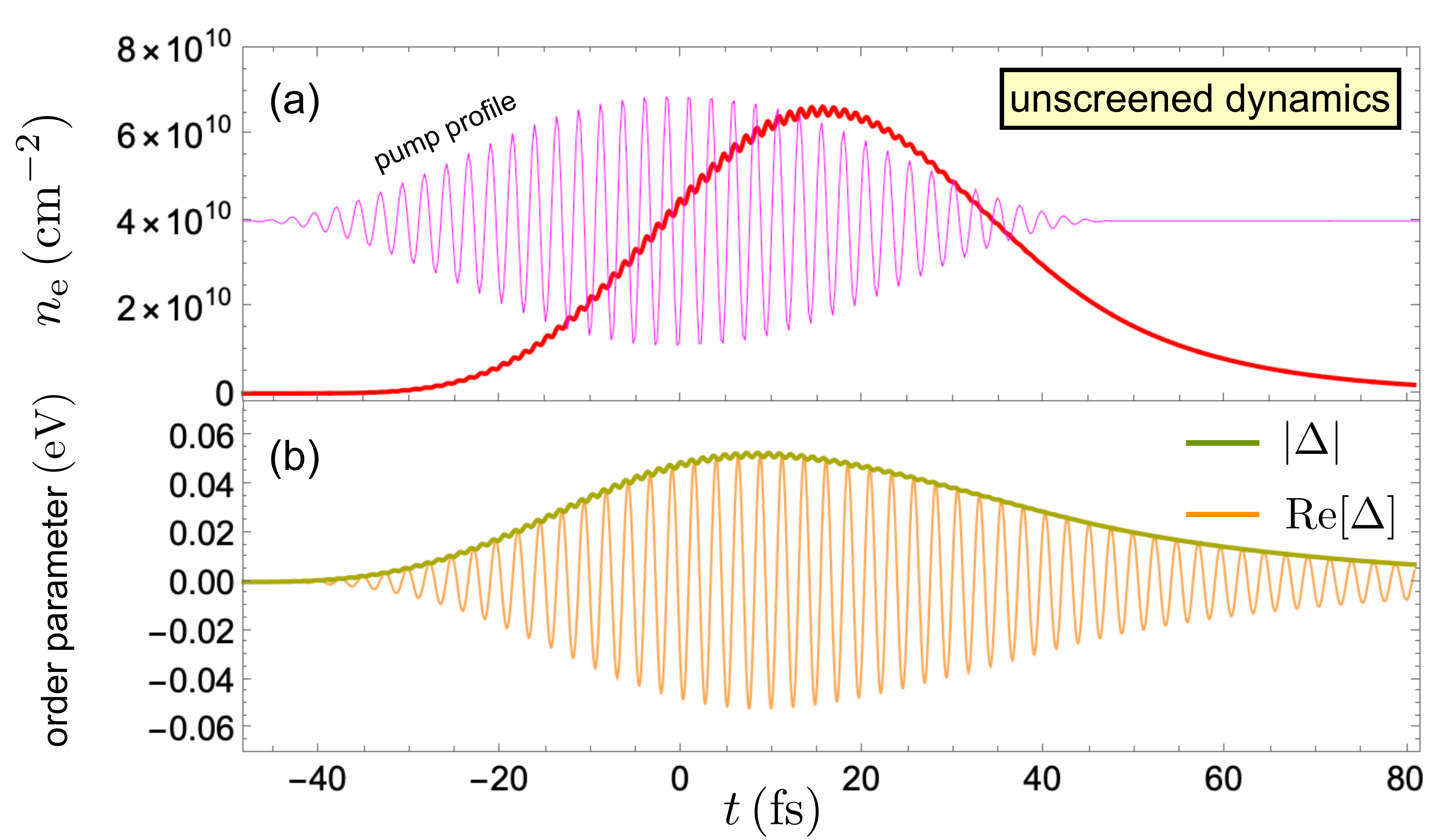}
\caption{Evolution of the conduction density $n_{\rm e}(t)$ at the K 
point (panel a) and order 
parameter $\D(t)$ (panel b) without screening effects. The  
profile of the pump field is displayed in panel a (magenta curve).}
\label{figtdhf}
 \end{figure}

\section{Unscreened dynamics: Robustness of the NEQ-EI phase}
\label{sec5}

\begin{figure}[tbp]
\includegraphics[width=0.45\textwidth]{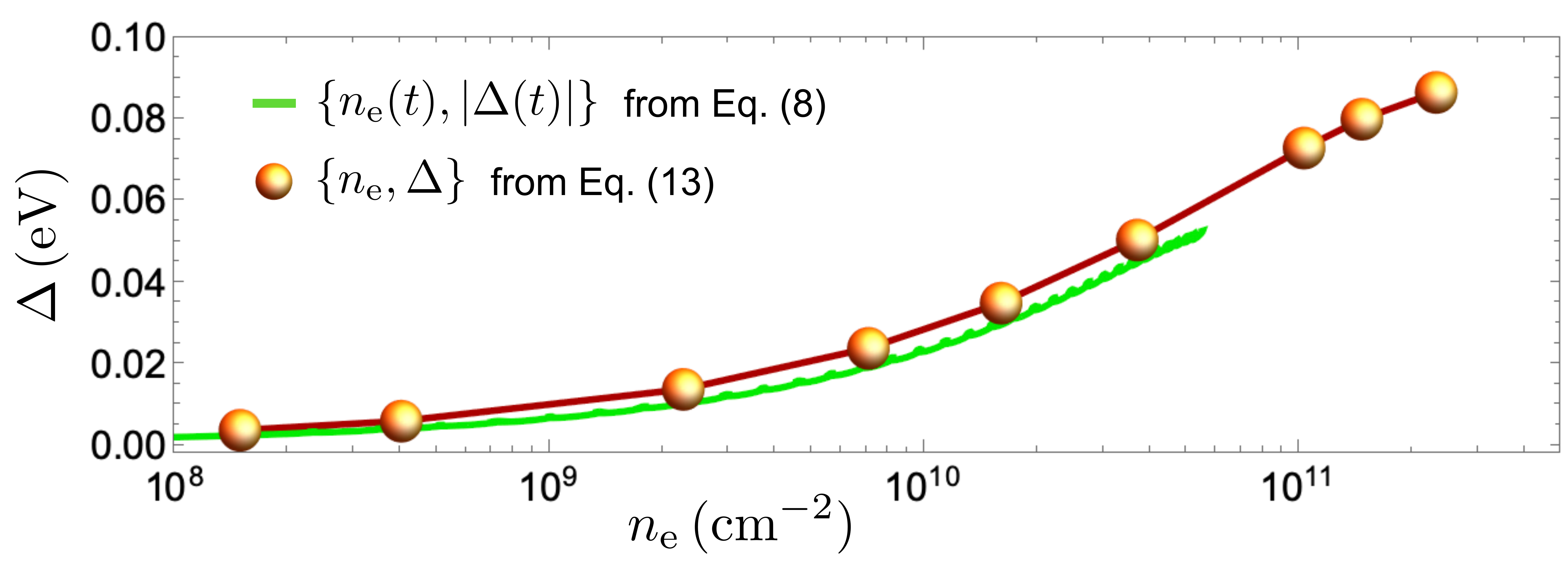}
\caption{Comparison between the self-consistent pairs $\{ n_{\rm e},\D \}$ 
of Fig.~(\ref{figphase}) (yellow bullets) and the time-dependent
pairs $\{ n_{\rm e}(t),\D(t) \}$ of 
Fig.~\ref{figtdhf} (green curve).}
\label{figad}
\end{figure}

\begin{figure*}[tbp]
\includegraphics[width=1.0\textwidth]{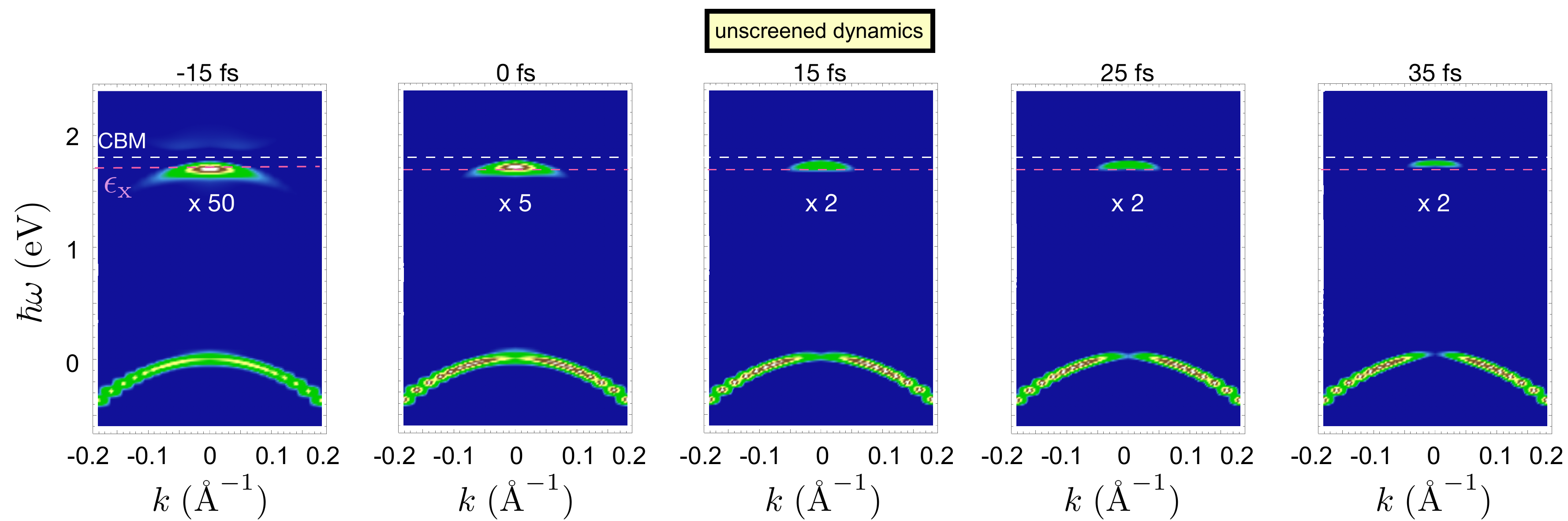}
\caption{Transient spectral function $A_{\blk}(\t,\w)$ evaluated without screening effects
at different delays. The intensity from the conduction band is 
multiplied by the factor $M=2S_{v}/S_{c}$ where $S_{\a}$ is the  
spectral weight of band $\a=v,c$ at $\blk=0$.
Energies are measured with respect to the 
the valence band maximum. Horizontal dashed lines at energies 
$\e_{g}$ (white) and $\e_{\mathrm{\rm x}}$ (red)  are drawn to guide the eye.}
\label{trarpesgamma}
\end{figure*} 
 
We first investigate the effects of intervalley scattering by 
neglecting screening effects, i.e., we propagate  Eq.~(\ref{eomtdhpsex}) 
by setting $W_{\blq}(t)=U_{\blq}$.
In Fig.~\ref{figtdhf} we plot the evolution of the excited 
density $n_{\rm e}(t)$ (panel a) and  order parameter $\D(t)$ (panel b) during and after pumping.
We see that $n_{\rm e}(t)$ reaches a maximum value about $20$~fs after the 
pump peak, and then it decays exponentially on the time-scale $\t_{\mathrm{K}}=15$~fs.
The intervalley scattering is also responsible 
for an exponential decay  of the amplitude of the order 
parameter $\D(t)$ on the time-scale $\sim 2 
\t_{\mathrm{K}}$. However, 
coherence is preserved since the order parameter continues to oscillate 
monochromatically.

In Fig.~\ref{figad} we  show the points $\{n_{\rm e},\D\}$
belonging to self-consistent solution of Fig.~\ref{figphase} together 
with  the parametric plot of the points $\{n_{\rm e}(t),|\D(t)|\}$ belonging to 
the temporal evolution of Fig.~\ref{figtdhf}.
For times $t\lesssim 20$~fs, i.e., before the conduction density $n_{\rm e}(t)$ 
reaches its maximum value, the system visits 
instantaneously all NEQ-EI states of the self-consistent solution.
This ``adiabatic'' behaviour is a consequence of the resonant pumping.
Indeed a weak photoexcitation can only create coherent 
excitons if  $\w_{P}=\e_{\rm x}<\e_{g}$  (quasi-particle states are accessible only for 
$\w_{P}>\e_{g}$) and therefore $N_{\mathrm{ex}}(t) =  N_{\mathrm{e}}(t)$, see 
Eq.~(\ref{bec}).

Due to the intervalley scattering the spectral function becomes 
time-dependent. In Fig.~\ref{trarpesgamma} we show $A_{\blk}(\t,\w)$ 
for different times $\t$. 
As expected, for small delays $A_{\blk}(\t,\w)$ displays the typical excitonic 
sideband inside the gap, lying exactly  $\e_{b}=0.1$~eV below the 
CBM.  This further corroborates the adiabatic scenario according to 
which  the NEQ-EI state
is generated during pumping, see inset of Fig.~\ref{figphase}.
More importantly, the excitonic sideband survives also at larger delays.
This implies that conduction electrons at K remain bound and 
excitons do not break into free e-h pair at K but only into 
free electrons at $\S$ and free holes at K. 
We conclude that an uncontaminated NEQ-EI phase, or equivalently a BEC state of coherent 
excitons, exists until all excitons break. 

\section{Screened Dynamics: Ultrafast melting of the NEQ-EI phase}
\label{sec6}

\begin{figure}[tbp]
    \includegraphics[width=0.45\textwidth]{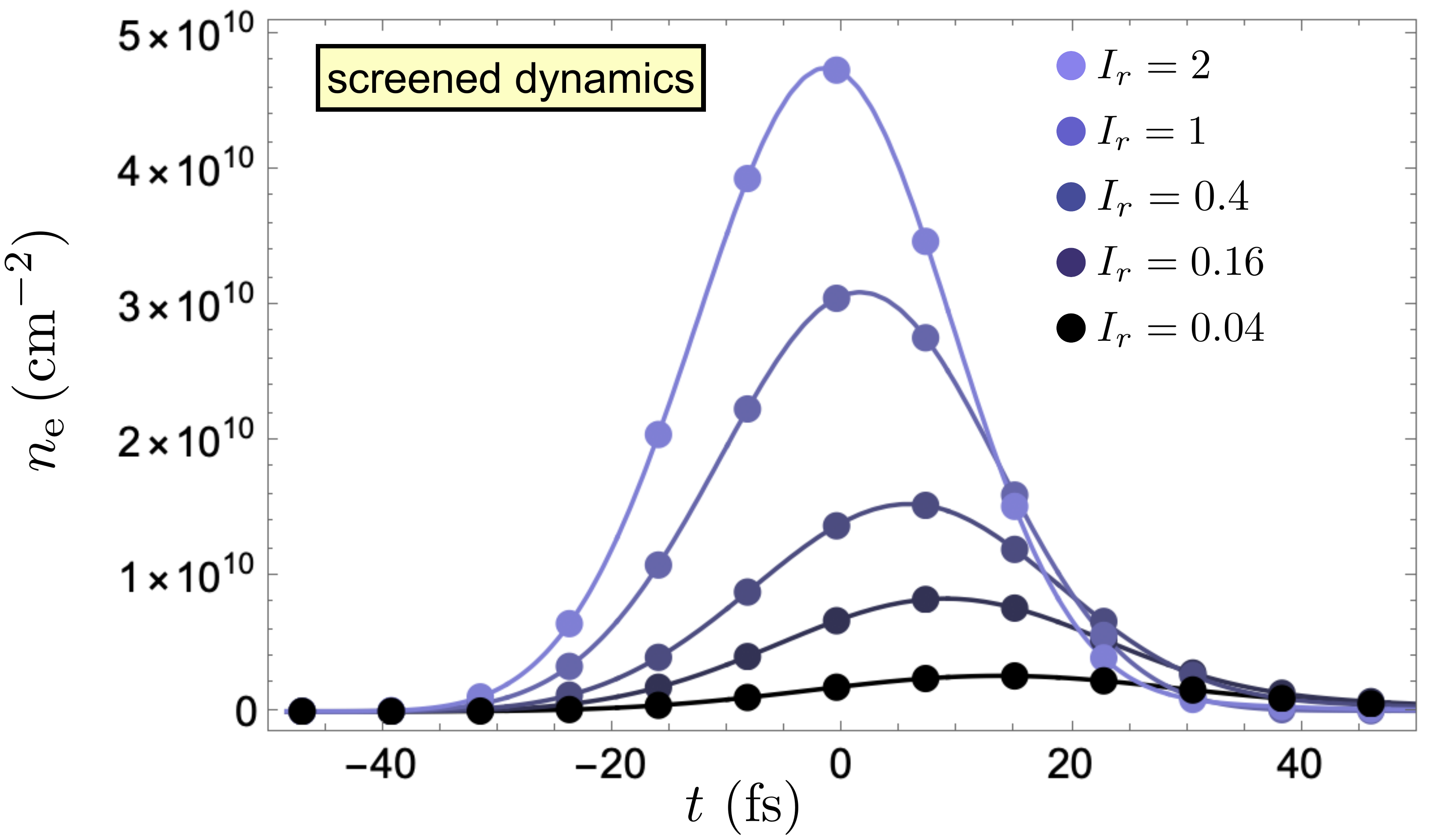}
 \caption{Temporal evolution of the excited density $n_{\mathrm{e}}(t)$  for 
 different pump intensities $I_{r}$ in the TD-HSEX approximation.}
 \label{intensity}
 \end{figure}

In this Section we discuss how the dynamics changes 
when screening is included. Before, however,
we assess the validity of the time-dependent plasma screening  
approximation in Eqs.~(\ref{hpsex}) and (\ref{xpsex}).
To this end we study
how $n_{\rm e}(t)$ varies as the intensity of the pump field 
increases. This issue has been experimentally investigated in 
Ref.~\cite{Puppin-PhD} and it has been found that the time at which $n_{\rm e}(t)$ 
is maximum approaches zero with increasing intensity.
In Fig.~\ref{intensity} we show $n_{\rm e}(t)$ for different relative 
intensities $I_{r}=(\W_{P}/\W_{0})^{2}$, with $\W_{0}=13$~meV.
The experimental trend is fairly reproduced. We emphasize that 
the screened dynamics is crucial for this 
agreement. In fact, simulations performed with unscreened interaction 
reveal that the of maximum of $n_{\rm e}(t)$ is independent of the 
pump intensity (not shown) .

\begin{figure}[tbp]
 \includegraphics[width=0.45\textwidth]{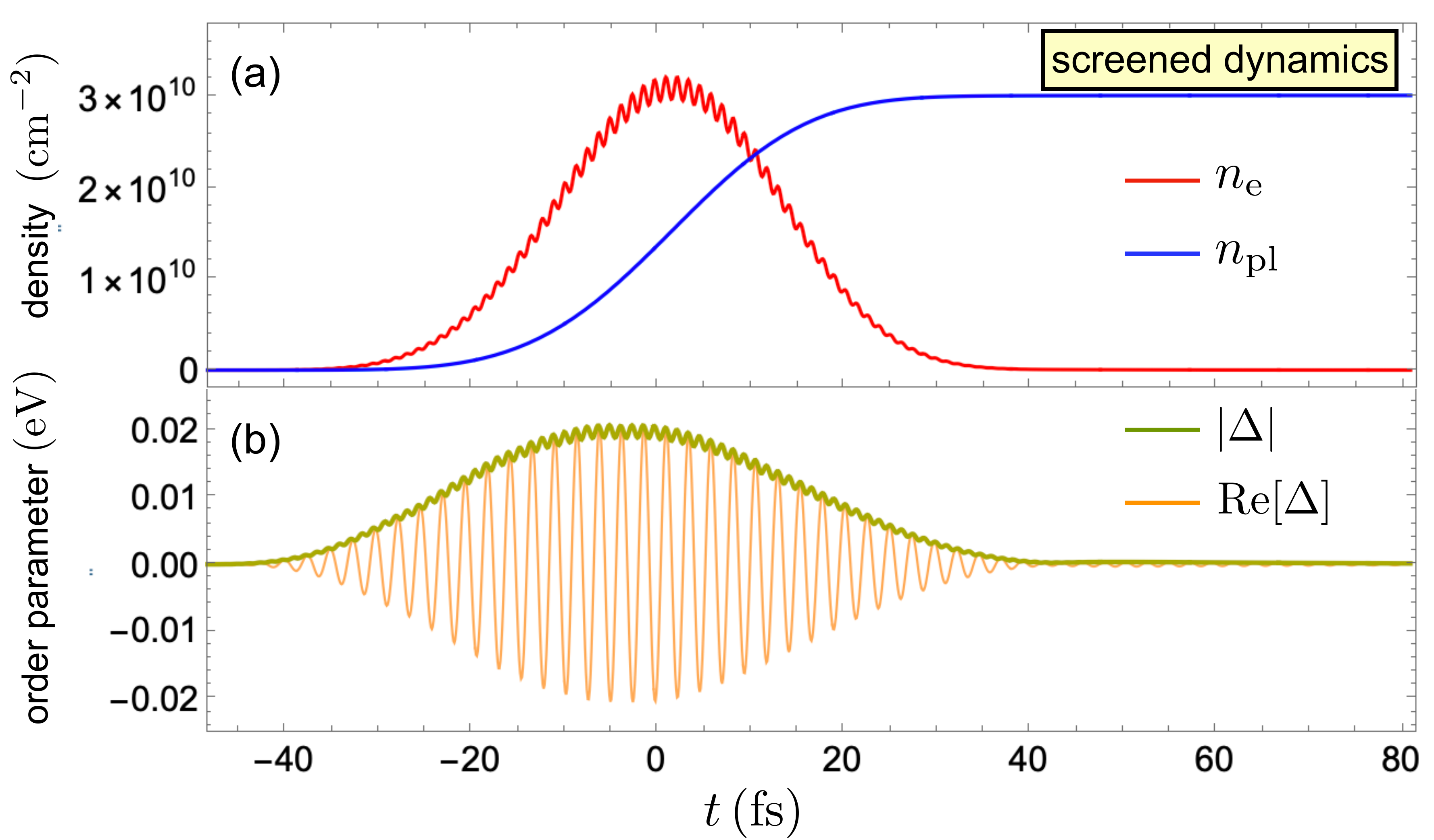}
 \caption{Temporal evolution of the excited density $n_{\mathrm{e}}(t)$ (red curve) and 
 screening density $n_{\mathrm{pl}}(t)$ (blue curve) (panel a) and order 
 parameter $\D(t)$ (panel b) in the TD-HSEX approximation.  }
 \label{tdhpsex}
 \end{figure}

In Fig.~\ref{tdhpsex} we show again the conduction density and the 
order parameter when both intevalley scattering and screening are 
taken into account. The pump pulse is the same as in 
Fig.~\ref{figtdhf}.
The maximum value of $n_{\rm e}(t)$ is twice 
smaller than in the unscreened dynamics and it occurs at time $t\approx 0$, 
i.e., when the pump pulse is maximum. After this time 
electrons rapidly migrate  to the $\S$ 
valley and at $t\approx 35 \div 40$~fs the migration is complete -- 
no excited carriers in the K valley. 
The migration generates  a plasma of 
free holes in the valence band and hence a nonvanishing 
$n_{\mathrm{pl}}$, see panel a (blue curve).
In the early transient  $t\lesssim -25$~fs the plasma density
is very small and screening is negligible. According to the findings 
of the previous Section the system is in 
pure NEQ-EI state.
As $n_{\mathrm{pl}}$ becomes sizable the electron-hole 
attraction is drastically reduced. The order parameter, 
see panel b,  reaches a maximum 
value already at $t\approx -5$~fs, and then 
decays at a much faster rate than $1/2\t_{\mathrm{K}}$ (the 
unscreened rate). To shed light on the physical scenario in 
this stage we calculate the transient spectral function.

\begin{figure*}[tbp]
\includegraphics[width=1.0\textwidth]{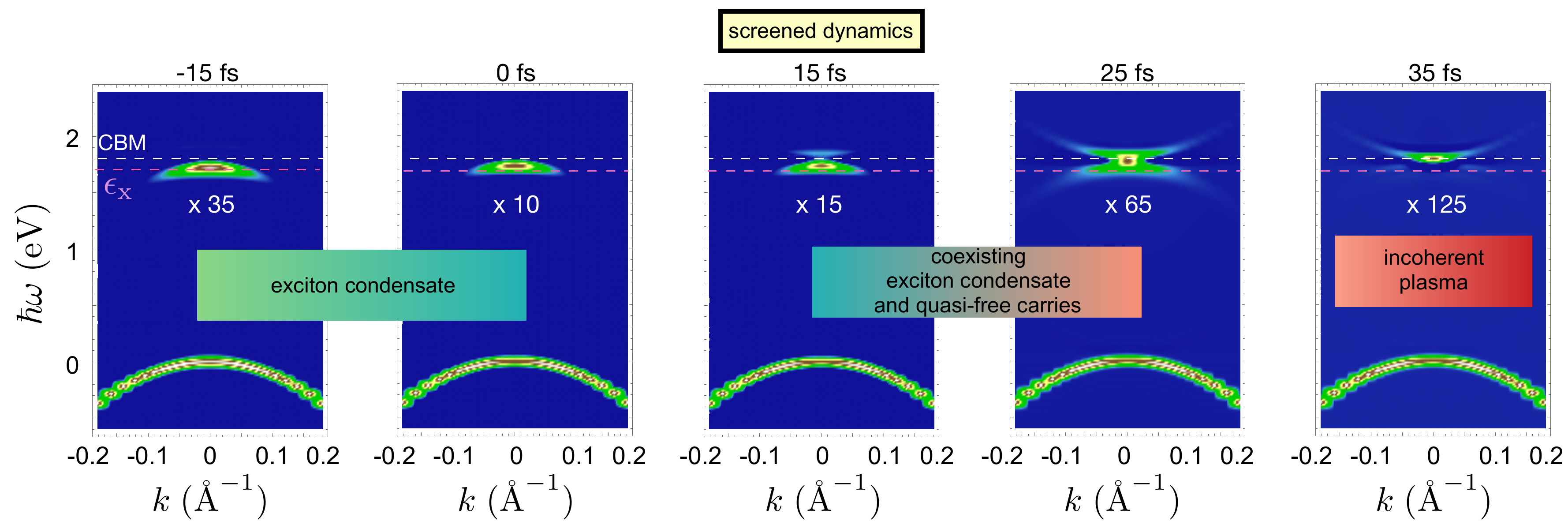}
\caption{Transient spectral function $A_{\blk}(\t,\w)$ evaluated with screening effects
at different delays. The intensity from the conduction band is 
multiplied by the factor $M=S_{v}/S_{c}$ where $S_{\a}$ is the  
spectral weight of band $\a=v,c$ at $\blk=0$.
Energies are measured with respect to the 
the valence band maximum. Horizontal dashed lines at energies 
$\e_{g}$ (white) and $\e_{\mathrm{\rm x}}$ (red)  are drawn to guide the eye.}
\label{trarpes}
\end{figure*} 

In Fig.~\ref{trarpes} we show $A_{\blk}(\t,\w)$ for the same 
delays $\t$ as in Fig.~\ref{trarpesgamma}.
For   $\t \lesssim 0$ the spectral function exhibits the typical 
feature of the NEQ-EI phase, with an excitonic replica of the valence 
band at energy $\e_{\mathrm{\rm x}}<\e_{g}$.
For these delays the unscreened and screened spectral functions 
are almost identical.
At delays $\t\gtrsim 15$~fs a spectral structure 
right above the CBM develops. 
The appearance of this second structure in the spectral function 
points to a clear picture:
excitons partially dissociate by breaking into \mbox{e-h} pairs at the 
K-point, with the free electrons occupying the empty 
levels around the CBM. Therefore excitons in the 
NEQ-EI phase and a plasma of free carriers coexist. As time increases the number of bound e-h 
pairs at the K-point becomes smaller and 
eventually the spectral weight is totally 
transferred to the conduction band.
At delays $t\gtrsim 35$~fs only the conduction band is visible, 
implying that all carries are free.
This incoherent regime lasts until the migration from K to $\S$ is 
completed, i.e., until time  $t\approx 40$~fs.

\section{Conclusions and outlooks}
\label{sec7}
 
We have studied the screened dynamics of the excitonic condensate 
forming in a bulk WSe$_{2}$ upon pumping in resonance with the 
lowest-energy exciton. 
Through the transient spectral function we have been able to observe 
the transition from an initial NEQ-EI  phase of coherent excitons 
to a final phase of incoherent 
e-h pairs. This transition is not abrupt as the two phases coexist.
The proposed theory relies on a general 
mechanism based on the interplay between intervalley scattering and 
plasma screening and the results agree with recent findings on the same 
system~\cite{Puppin-PhD}. In fact, neglecting the renormalization of 
the effective e-h attraction the excitons at the K point would not break 
into e-h pairs at the same point and hence no signal from the 
conduction band would be detected at K. Furthermore, the maximum value of 
the density in the conduction band occurs at a delay $\t$  which
approaches zero with increasing the intensity of the pump pulse.

The screening due to quasi-free holes arising from the intervalley 
scattering is responsible 
for an ultrafast melting of the NEQ-EI phase. Although this mechanism 
has been highlighted in WSe$_{2}$ it is likely to occur
in other indirect gap semiconductors as the only 
condition to meet is that electrons migrating from a local valley 
of the conduction band to the global CBM do not bounce back.

We have presented results  based on a 2D two-band model. However 
the underlying theory can be implemented in available 
first-principles time-dependent codes~\cite{Sangalli2019} to account explicitely for the 
spin-orbit interaction, the 
$\S$ valley degrees of freedom as well as 
the phonon-induced intervalley scattering.
The effective nonequilibrium e-h interaction is indeed screened adiabatically, hence 
retardation effects are discarded.
In this work the interaction has been screened at the RPA level 
using the static 2D Lindhard function 
calculated at the density of the hole plasma
-- excitonic screening is negligible in the dilute 
limit~\cite{PhysRevB.102.085203}. In first-principles implementations
the nonequilibrium response function should instead be built taking into account the 
real band structure of the material. 

 \vspace{1cm}
 
{\it Acknowledgements}
We acknowledge useful discussions with Andrea Marini and Davide Sangalli.
We also acknowledge funding from MIUR PRIN Grant No. 20173B72NB and 
from INFN20-TIME2QUEST project.
G.S. acknowledges Tor Vergata University
for financial support through the Beyond Borders Project ULEXIEX.

\appendix
\section{Self-consistent screening in the NEQ-EI superfluid} 

\begin{figure}[tbp]
 \includegraphics[width=0.4\textwidth]{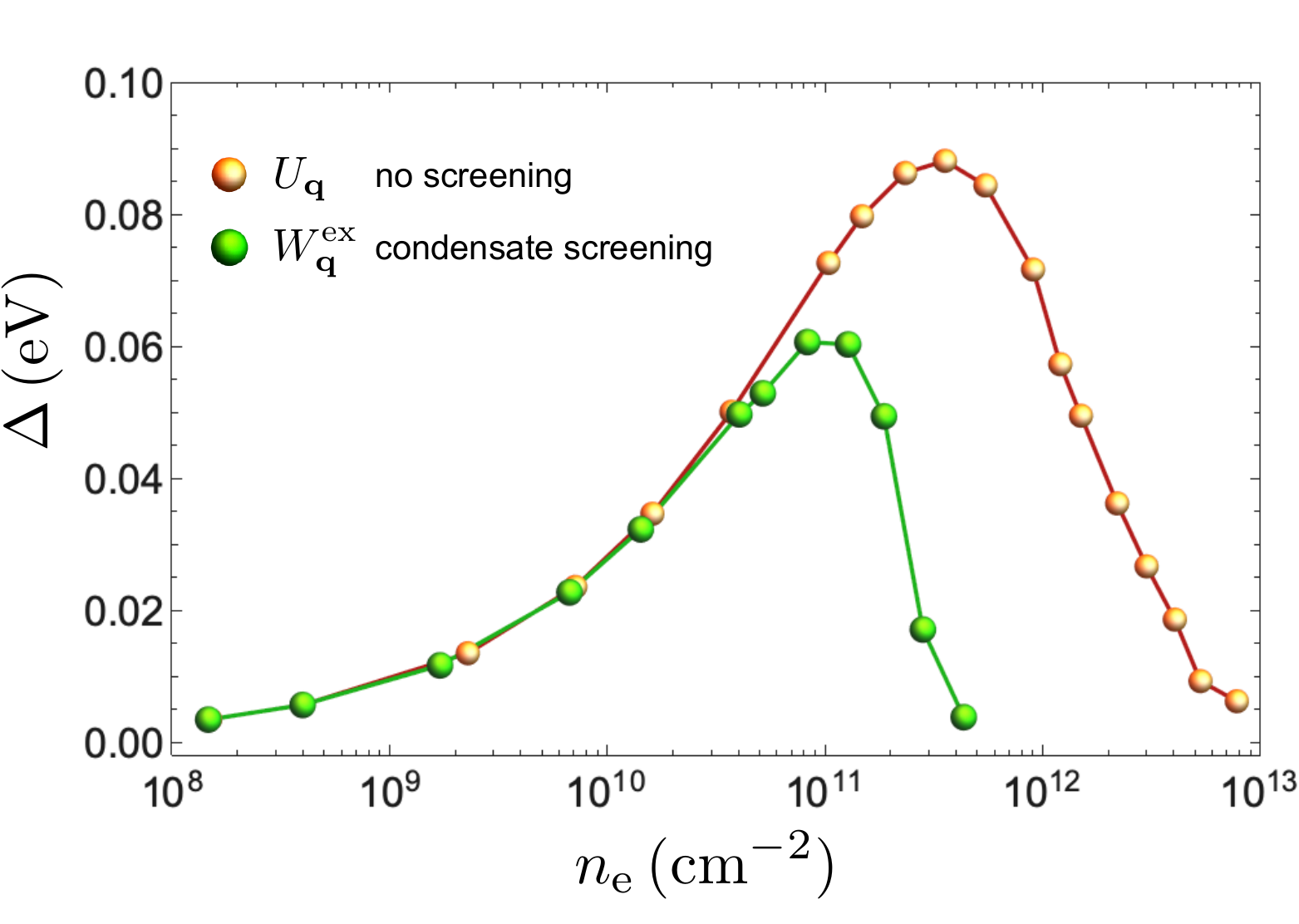}
 \caption{Order parameter with and without self-consistent 
 excitonic screening versus electronic density $n_{\rm e}$.}
 \label{figphase2}
 \end{figure}

In the pure NEQ-EI state, i.e., the solution of the secular problem in 
Eq.~(\ref{hhsex}), there is a finite density of conduction electrons 
$n_{\mathrm{e}}$ and valence holes $n_{\mathrm{h}}=n_{\mathrm{e}}$.
It is therefore natural to ask whether in $\mathrm{WSe}_{2}$ bulk 
these excited carriers are capable to screen the e-h attraction  such to
disrupt the superfluid state and, eventually, to restore the normal phase.
In a recent work~\cite{PhysRevB.102.085203} we have shown that
the  Coulomb 
repulsion $W^{\mathrm{ex}}$ {\it screened} by the excitonic condensate is given by
\be
W^{\mathrm{ex}}_{\blq}=\frac{U_{\blq}}{1-U_{\blq}
(\chi^{ \blq}_{vv} +\chi^{\blq}_{cc} +2\chi^{ \blq}_{vc} )},
\label{hsex}
\ee
with the excitonic response function given by
\bea
\chi^{\blq}_{\a \b}
=\frac{2}{\callN}\sum_{\blk} \left[
\frac{ \varphi^{+}_{\a \blk +\blq}  \varphi^{+}_{\b \blk +\blq} 
 \varphi^{-}_{\a \blk }  \varphi^{-}_{\b \blk}}{-(e^{+}_{ \blk 
 +\blq}-e^{-}_{ \blk })+i\eta} \right. \nonumber \\
 -
 \left. \frac{ \varphi^{-}_{\a \blk +\blq}  \varphi^{-}_{\b \blk +\blq} 
 \varphi^{+}_{\a \blk }  \varphi^{+}_{\b \blk}}{-(e^{-}_{ \blk 
 +\blq}-e^{+}_{ \blk })+i\eta}
\right].
\label{chiw}
\eea
Therefore the NEQ-EI state in the presence of the self-generated 
screening must be obtained by solving  Eq.~(\ref{hhsex}) by replacing 
$U_{\blq}\to W^{\mathrm{ex}}_{\blq}$. The resulting order parameter
is displayed in Fig.~\ref{figphase2}. In order to highlight the 
impact of the excitonic screening we also show the points of 
Fig.~\ref{figphase} resulting from a self-consistent calculation with 
the bare interaction $U_{\blq}$.
We observe that  for low densities $n_{\mathrm{e}} \lesssim  
\times 10^{11}\mathrm{cm}^{-2}$  screening  effects are 
almost irrelevant.
We recall that in the low-density regime  
{\it all} excited carriers participate to the creation of 
zero-momentum excitons, i.e. $N_{\mathrm{e}}=N_{\mathrm{ex}}$ (see 
Appendix B). The absence of free carriers and the fact that 
excitons are neutral bound states explain why in the dilute 
limit the screening efficiency is so scarce.

This result justifies the neglect of condensate screening
in the numerical calculation of Figs.~\ref{figtdhf} and ~\ref{tdhpsex}, where the maximum 
excited density is $n_{\rm e}\approx 6 \times 10^{10}\mathrm{cm}^{-2}$ and 
$n_{\rm e}\approx 3 \times 10^{10}\mathrm{cm}^{-2}$ respectively.

\section{Proof of Eq.~(\ref{bec})} 

The NEQ-EI many-body state corresponding to the solution of  the secular problem in 
Eq.~(\ref{hhsex}) can be written as
\be
|\Psi_{\mathrm{x}}\ket = \prod_{\blp \s}(\varphi^{-}_{v \blp } 
\hat{f}^{\dag}_{v\blp \s}  + \varphi^{-}_{c \blp } 
\hat{f}^{\dag}_{c\blp \s})|0\ket \, ,
\ee
where  $\hat{f}_{\m \blp \s} ^{(\dag)}$ annihilates (creates)
an electron of momentum $\blk$ with spin $\s$ in band $\m =c,v$, 
and $|0\ket$ is the electron vacuum.
Let us consider the creation operator of an exciton with momentum $\blq$ 
\be
\hat{b}^{\dag}_{\blq}=\sum_{\blk\s} Y^{\blq}_{\blk} 
\hat{f}^{\dag}_{c\blk \s}\hat{f}_{v\blk \s} ,
\ee
where $ Y^{\blq}_{\blk} $ is the excitonic wavefunction in momentum 
space, and define the total number of excitons with momentum $\blq$  as
\be
N_{\mathrm{ex} ,\blq}\equiv \bra \Psi_{\mathrm{x}} | 
\hat{b}^{\dag}_{\blq} 
\hat{b}_{\blq}| \Psi_{\mathrm{x}} \ket .
\ee
It is  matter of simple algebra to show that
\begin{align}
N_{\mathrm{ex}, \blq}&=  2\d_{\blq \mathbf{0}} \left[ 
\sum_{\blk}(Y^{\mathbf{0}}_{\blk}  \r^{cc}_{\blk} )^{2} + \sum_{\blk 
\neq \blk' }  Y^{\mathbf{0}}_{\blk}  Y^{\mathbf{0}}_{\blk'} \r^{cv}_{\blk}
\r^{vc}_{\blk}  \right] \nonumber \\
 &+2 (1-\d_{\blq \mathbf{0}})\sum_{\blk} (Y^{\blq}_{\blk}  
 \varphi^{-}_{c \blk } \varphi^{-}_{c \blk +\blq } )^{2} \, .
 \label{nex1}
\end{align}
In the low-density limit the eigenvectors $\varphi^{-}_{\a \blk }$ 
can be written in terms of the zero-momentum excitonic 
wavefunction $Y_{\blq} \equiv  Y^{\mathbf{0}}_{\blk}$ using 
Eq.~(\ref{linearr}). Taking into account the normalization condition 
$ \sum_{\blk} |Y_{\blk}|^{2} =1$ we find
\be
N_{\mathrm{ex}, \blq}= \d_{\blq \mathbf{0}}N_{\mathrm{e}} + (1-\d_{\blq 
\mathbf{0}}) N_{\mathrm{e}}^{2} \sum_{\blk}  (Y^{\blq}_{\blk}  
Y_{\blk} Y_{\blk +\blq} )^{2} \, .
\ee
It  follows that 
$N_{\mathrm{ex}, \mathbf{0}} \equiv N_{\mathrm{ex}} = N_{\mathrm{e}} $
diverges in the thermodynamic limit for any finite electron density 
$n_{\rm e}$. The number of nonvanishing momentum excitons 
$N_{\mathrm{ex} \blq \neq  \mathbf{0} }$ is instead of order 
$\callO(\callN^{0})$.

\section{Evaluation of the transient spectral function} 

The tr-ARPES signal is proportional to the 
number of electrons $N_{\blk}(\w)$  with 
energy $\w$ and parallel momentum $\blk$
ejected by a probe pulse. 
For an arbitrary probe pulse of temporal profile $e(t)$ we have~\cite{PSMS.2016,Freericks_PhysRevLett.102.136401}
\be
N_{\blk}(\w)=2\sum_{\m\n}\int dt_{1} dt_{2}\;\Re\left[
\S^{\m\n,\rm 
R}_{\blk,\w}(t_{1},t_{2})G^{\n\m,<}_{\blk}(t_{2},t_{1})\right].
\label{Nkw}
\ee
Here $G^{\n\m,<}_{\blk}(t_{2},t_{1})$ is the (spin-independent) lesser Green's function
defined as~\cite{svl-book}
\be
G^{\n\m,<}_{\blk}(t_{2},t_{1})=\bra  \hat{f}^{\dag}_{\m \blk \s 
}(t_{1})  \hat{f}_{\n \blk \s
}(t_{2})   \ket \, ,
\ee
with time dependent  operators $\hat{f}^{(\dag)}_{\m \blk \s }(t)$ 
in the  Heisenberg picture.
The ionization self-energy  reads~\cite{PUvLS.2015}
\be
\S^{\m\n,\rm R}_{\blk,\w}(t_{1},t_{2})=-i\th(t_{1}-t_{2})
d_{\blk,\w}^{\m} e(t_{1})e^{-i\w(t_{1}-t_{2})}
d_{\blk,\w}^{\n} e(t_{2})
\ee
where $d_{\blk,\w}^{\m}$ is the dipole matrix element between a state 
in band $\m$ and momentum $\blk$ and a continuum time-reversed LEED state of energy  $\w$ and parallel momentum 
$\blk$. 

From Eq.~(\ref{Nkw}) we see that $N_{\blk}(\w)$ is a complicated 
two-times convolution of $G^{<}$.
However a simpler and more transparent expression can be obtained if 
the probe pulse has a 
duration $t_{p}$ much longer than
the typical electronic timescale
and  a frequency $\w_{p}$ large enough to resolve  
the desired removal energies.
In this case one can shown that Eq.~(\ref{Nkw}) reduces to~\cite{PBS.2020}
\be 
N_{\blk}(\w)\propto A_{\blk}(\t,\w-\w_{p})
\label{trarpessignal}
\ee
where $\t$ is the time at which the probe impinges the system and  
transient spectral function 
\bea
A_{\blk}(\t,\w)=-i\int_{\t-t_{p}/2}^{\t+t_{p}/2}  d\bar{t}\,e^{i\w 
\bar{t}} \, \mathrm{Tr} 
[G^{<}_{\blk}(\t-\bar{t},\t+\bar{t})].
\label{trARPESAA}
\eea
In bulk  $\mathrm{WSe}_{2}$ excited by a VIS pump the density matrix varies in time no slower 
than a timescale of $\sim 2$~fs, given by the inverse of the lowest exciton energy 
$\e_{\mathrm{x}}$.
The approximated expression in 
Eq.~(\ref{trARPESAA}) can be safely used if the XUV probe has central frequency $ \w_{p}\approx 
20$~eV and total duration $t_{p}\approx 70$~fs (FWHM = 35~fs) .

To obtain the lesser Green's function $G^{<}$ from the numerical solution
of Eq.~(\ref{eomtdhpsex}) we use the relation~\cite{PhysRevB.34.6933}
\be
G_{\blk}^{<}(t,t')=-G_{\blk}^{R}(t',t')\r_{\blk}(t)
+\r_{\blk}(t)G_{\blk}^{\rm A}(t,t')\, .
\ee
Here the retarded Green's function reads~\cite{LPUvLS.2014} 
\be
G^{\rm R}_{\blk}(t,t')=-i\th(t,t')\callT\left\{
e^{-i\int_{t'}^{t}d\bar{t}[h_{\mathrm{HSEX}, \blk}(t) 
-i\G/2]}\right\} \, ,
\ee
where $\callT$ is the time-ordering operator and  $G^{\rm R}_{\blk}(t,t')=[G^{\rm A}_{\blk}(t',t)]^{\ast}$.


\begin{thebibliography}{75}
\expandafter\ifx\csname natexlab\endcsname\relax\def\natexlab#1{#1}\fi
\expandafter\ifx\csname bibnamefont\endcsname\relax
  \def\bibnamefont#1{#1}\fi
\expandafter\ifx\csname bibfnamefont\endcsname\relax
  \def\bibfnamefont#1{#1}\fi
\expandafter\ifx\csname citenamefont\endcsname\relax
  \def\citenamefont#1{#1}\fi
\expandafter\ifx\csname url\endcsname\relax
  \def\url#1{\texttt{#1}}\fi
\expandafter\ifx\csname urlprefix\endcsname\relax\def\urlprefix{URL }\fi
\providecommand{\bibinfo}[2]{#2}
\providecommand{\eprint}[2][]{\url{#2}}

\bibitem[{\citenamefont{Keldysh}(1972)}]{keldysh1972problems}
\bibinfo{author}{\bibfnamefont{L.}~\bibnamefont{Keldysh}},
  \emph{\bibinfo{title}{Problems of theoretical physics}}
  (\bibinfo{year}{1972}).

\bibitem[{\citenamefont{Keldysh and Kozlov}(1968)}]{Keldysh-Kozlov_JETP1968}
\bibinfo{author}{\bibfnamefont{L.~V.} \bibnamefont{Keldysh}} \bibnamefont{and}
  \bibinfo{author}{\bibfnamefont{A.~N.} \bibnamefont{Kozlov}},
  \bibinfo{journal}{JETP} \textbf{\bibinfo{volume}{27}}, \bibinfo{pages}{521}
  (\bibinfo{year}{1968}).

\bibitem[{\citenamefont{{Moskalenko} and {Snoke}}(2000)}]{2000bceb}
\bibinfo{author}{\bibfnamefont{S.~A.} \bibnamefont{{Moskalenko}}}
  \bibnamefont{and} \bibinfo{author}{\bibfnamefont{D.~W.}
  \bibnamefont{{Snoke}}}, \emph{\bibinfo{title}{Bose-Einstein Condensation of
  Excitons and Biexcitons}} (\bibinfo{year}{2000}).

\bibitem[{\citenamefont{Combescot and Shiau}(2016)}]{combescot2016excitons}
\bibinfo{author}{\bibfnamefont{M.}~\bibnamefont{Combescot}} \bibnamefont{and}
  \bibinfo{author}{\bibfnamefont{S.}~\bibnamefont{Shiau}},
  \emph{\bibinfo{title}{Excitons and Cooper Pairs: Two Composite Bosons in
  Many-body Physics}}, Oxford graduate texts (\bibinfo{publisher}{Oxford
  University Press}, \bibinfo{year}{2016}), ISBN \bibinfo{isbn}{9780198753735},
  \urlprefix\url{https://books.google.it/books?id=YzAiCwAAQBAJ}.

\bibitem[{\citenamefont{Schmitt-Rink et~al.}(1988)\citenamefont{Schmitt-Rink,
  Chemla, and Haug}}]{Schmitt-Rink_PhysRevB.37.941}
\bibinfo{author}{\bibfnamefont{S.}~\bibnamefont{Schmitt-Rink}},
  \bibinfo{author}{\bibfnamefont{D.~S.} \bibnamefont{Chemla}},
  \bibnamefont{and} \bibinfo{author}{\bibfnamefont{H.}~\bibnamefont{Haug}},
  \bibinfo{journal}{Phys. Rev. B} \textbf{\bibinfo{volume}{37}},
  \bibinfo{pages}{941} (\bibinfo{year}{1988}),
  \urlprefix\url{https://link.aps.org/doi/10.1103/PhysRevB.37.941}.

\bibitem[{\citenamefont{Kuklinski and Mukamel}(1990)}]{kuklinski1990}
\bibinfo{author}{\bibfnamefont{J.~R.} \bibnamefont{Kuklinski}}
  \bibnamefont{and} \bibinfo{author}{\bibfnamefont{S.}~\bibnamefont{Mukamel}},
  \bibinfo{journal}{Physical Review B} \textbf{\bibinfo{volume}{42}},
  \bibinfo{pages}{2959} (\bibinfo{year}{1990}).

\bibitem[{\citenamefont{Glutsch and
  Zimmermann}(1992)}]{Glutsch_PhysRevB.45.5857}
\bibinfo{author}{\bibfnamefont{S.}~\bibnamefont{Glutsch}} \bibnamefont{and}
  \bibinfo{author}{\bibfnamefont{R.}~\bibnamefont{Zimmermann}},
  \bibinfo{journal}{Phys. Rev. B} \textbf{\bibinfo{volume}{45}},
  \bibinfo{pages}{5857} (\bibinfo{year}{1992}),
  \urlprefix\url{https://link.aps.org/doi/10.1103/PhysRevB.45.5857}.

\bibitem[{\citenamefont{Littlewood and Zhu}(1996)}]{littlewood1996}
\bibinfo{author}{\bibfnamefont{P.}~\bibnamefont{Littlewood}} \bibnamefont{and}
  \bibinfo{author}{\bibfnamefont{X.}~\bibnamefont{Zhu}},
  \bibinfo{journal}{Physica Scripta} \textbf{\bibinfo{volume}{1996}},
  \bibinfo{pages}{56} (\bibinfo{year}{1996}).

\bibitem[{\citenamefont{{\"O}streich and
  Sch{\"o}nhammer}(1993)}]{Ostreich_1993}
\bibinfo{author}{\bibfnamefont{T.}~\bibnamefont{{\"O}streich}}
  \bibnamefont{and}
  \bibinfo{author}{\bibfnamefont{K.}~\bibnamefont{Sch{\"o}nhammer}},
  \bibinfo{journal}{Zeitschrift f{\"u}r Physik B Condensed Matter}
  \textbf{\bibinfo{volume}{91}}, \bibinfo{pages}{189} (\bibinfo{year}{1993}),
  \urlprefix\url{https://doi.org/10.1007/BF01315235}.

\bibitem[{\citenamefont{Hannewald et~al.}(2000)\citenamefont{Hannewald,
  Glutsch, and Bechstedt}}]{Hannewald-Bechstedt_2000}
\bibinfo{author}{\bibfnamefont{K.}~\bibnamefont{Hannewald}},
  \bibinfo{author}{\bibfnamefont{S.}~\bibnamefont{Glutsch}}, \bibnamefont{and}
  \bibinfo{author}{\bibfnamefont{F.}~\bibnamefont{Bechstedt}},
  \bibinfo{journal}{Journal of Physics: Condensed Matter}
  \textbf{\bibinfo{volume}{13}}, \bibinfo{pages}{275} (\bibinfo{year}{2000}),
  \urlprefix\url{https://doi.org/10.1088%2F0953-8984%2F13%2F2%2F305}.

\bibitem[{\citenamefont{Glutsch et~al.}(1992)\citenamefont{Glutsch, Bechstedt,
  and Zimmermann}}]{GLUTSCH1992}
\bibinfo{author}{\bibfnamefont{S.}~\bibnamefont{Glutsch}},
  \bibinfo{author}{\bibfnamefont{F.}~\bibnamefont{Bechstedt}},
  \bibnamefont{and}
  \bibinfo{author}{\bibfnamefont{R.}~\bibnamefont{Zimmermann}},
  \bibinfo{journal}{physica status solidi (b)} \textbf{\bibinfo{volume}{172}},
  \bibinfo{pages}{357} (\bibinfo{year}{1992}),
  \eprint{https://onlinelibrary.wiley.com/doi/pdf/10.1002/pssb.2221720131},
  \urlprefix\url{https://onlinelibrary.wiley.com/doi/abs/10.1002/pssb.2221720131}.

\bibitem[{\citenamefont{Perfetto et~al.}(2019)\citenamefont{Perfetto, Sangalli,
  Marini, and Stefanucci}}]{PSMS.2019}
\bibinfo{author}{\bibfnamefont{E.}~\bibnamefont{Perfetto}},
  \bibinfo{author}{\bibfnamefont{D.}~\bibnamefont{Sangalli}},
  \bibinfo{author}{\bibfnamefont{A.}~\bibnamefont{Marini}}, \bibnamefont{and}
  \bibinfo{author}{\bibfnamefont{G.}~\bibnamefont{Stefanucci}},
  \bibinfo{journal}{Phys. Rev. Materials} \textbf{\bibinfo{volume}{3}},
  \bibinfo{pages}{124601} (\bibinfo{year}{2019}),
  \urlprefix\url{https://link.aps.org/doi/10.1103/PhysRevMaterials.3.124601}.

\bibitem[{\citenamefont{Szyma\ifmmode~\acute{n}\else \'{n}\fi{}ska
  et~al.}(2006)\citenamefont{Szyma\ifmmode~\acute{n}\else \'{n}\fi{}ska,
  Keeling, and Littlewood}}]{SzymaPRL2006}
\bibinfo{author}{\bibfnamefont{M.~H.} \bibnamefont{Szyma\ifmmode~\acute{n}\else
  \'{n}\fi{}ska}}, \bibinfo{author}{\bibfnamefont{J.}~\bibnamefont{Keeling}},
  \bibnamefont{and} \bibinfo{author}{\bibfnamefont{P.~B.}
  \bibnamefont{Littlewood}}, \bibinfo{journal}{Phys. Rev. Lett.}
  \textbf{\bibinfo{volume}{96}}, \bibinfo{pages}{230602}
  (\bibinfo{year}{2006}),
  \urlprefix\url{https://link.aps.org/doi/10.1103/PhysRevLett.96.230602}.

\bibitem[{\citenamefont{Hanai et~al.}(2016)\citenamefont{Hanai, Littlewood, and
  Ohashi}}]{Hanai2016}
\bibinfo{author}{\bibfnamefont{R.}~\bibnamefont{Hanai}},
  \bibinfo{author}{\bibfnamefont{P.~B.} \bibnamefont{Littlewood}},
  \bibnamefont{and} \bibinfo{author}{\bibfnamefont{Y.}~\bibnamefont{Ohashi}},
  \bibinfo{journal}{Journal of Low Temperature Physics}
  \textbf{\bibinfo{volume}{183}}, \bibinfo{pages}{127} (\bibinfo{year}{2016}),
  \urlprefix\url{https://doi.org/10.1007/s10909-016-1552-6}.

\bibitem[{\citenamefont{Hanai et~al.}(2017)\citenamefont{Hanai, Littlewood, and
  Ohashi}}]{HanaiPRB2017}
\bibinfo{author}{\bibfnamefont{R.}~\bibnamefont{Hanai}},
  \bibinfo{author}{\bibfnamefont{P.~B.} \bibnamefont{Littlewood}},
  \bibnamefont{and} \bibinfo{author}{\bibfnamefont{Y.}~\bibnamefont{Ohashi}},
  \bibinfo{journal}{Phys. Rev. B} \textbf{\bibinfo{volume}{96}},
  \bibinfo{pages}{125206} (\bibinfo{year}{2017}),
  \urlprefix\url{https://link.aps.org/doi/10.1103/PhysRevB.96.125206}.

\bibitem[{\citenamefont{Triola et~al.}(2017)\citenamefont{Triola, Pertsova,
  Markiewicz, and Balatsky}}]{TriolaPRB2017}
\bibinfo{author}{\bibfnamefont{C.}~\bibnamefont{Triola}},
  \bibinfo{author}{\bibfnamefont{A.}~\bibnamefont{Pertsova}},
  \bibinfo{author}{\bibfnamefont{R.~S.} \bibnamefont{Markiewicz}},
  \bibnamefont{and} \bibinfo{author}{\bibfnamefont{A.~V.}
  \bibnamefont{Balatsky}}, \bibinfo{journal}{Phys. Rev. B}
  \textbf{\bibinfo{volume}{95}}, \bibinfo{pages}{205410}
  (\bibinfo{year}{2017}),
  \urlprefix\url{https://link.aps.org/doi/10.1103/PhysRevB.95.205410}.

\bibitem[{\citenamefont{Hanai et~al.}(2018)\citenamefont{Hanai, Littlewood, and
  Ohashi}}]{Hanai2018}
\bibinfo{author}{\bibfnamefont{R.}~\bibnamefont{Hanai}},
  \bibinfo{author}{\bibfnamefont{P.~B.} \bibnamefont{Littlewood}},
  \bibnamefont{and} \bibinfo{author}{\bibfnamefont{Y.}~\bibnamefont{Ohashi}},
  \bibinfo{journal}{Phys. Rev. B} \textbf{\bibinfo{volume}{97}},
  \bibinfo{pages}{245302} (\bibinfo{year}{2018}),
  \urlprefix\url{https://link.aps.org/doi/10.1103/PhysRevB.97.245302}.

\bibitem[{\citenamefont{Pertsova and Balatsky}(2018)}]{PertsovaPRB2018}
\bibinfo{author}{\bibfnamefont{A.}~\bibnamefont{Pertsova}} \bibnamefont{and}
  \bibinfo{author}{\bibfnamefont{A.~V.} \bibnamefont{Balatsky}},
  \bibinfo{journal}{Phys. Rev. B} \textbf{\bibinfo{volume}{97}},
  \bibinfo{pages}{075109} (\bibinfo{year}{2018}),
  \urlprefix\url{https://link.aps.org/doi/10.1103/PhysRevB.97.075109}.

\bibitem[{\citenamefont{Becker et~al.}(2019)\citenamefont{Becker, Fehske, and
  Phan}}]{Becker_PhysRevB.99.035304}
\bibinfo{author}{\bibfnamefont{K.~W.} \bibnamefont{Becker}},
  \bibinfo{author}{\bibfnamefont{H.}~\bibnamefont{Fehske}}, \bibnamefont{and}
  \bibinfo{author}{\bibfnamefont{V.-N.} \bibnamefont{Phan}},
  \bibinfo{journal}{Phys. Rev. B} \textbf{\bibinfo{volume}{99}},
  \bibinfo{pages}{035304} (\bibinfo{year}{2019}),
  \urlprefix\url{https://link.aps.org/doi/10.1103/PhysRevB.99.035304}.

\bibitem[{\citenamefont{Yamaguchi et~al.}(2013)\citenamefont{Yamaguchi, Kamide,
  Nii, Ogawa, and Yamamoto}}]{Yamaguchi_PhysRevLett.111.026404}
\bibinfo{author}{\bibfnamefont{M.}~\bibnamefont{Yamaguchi}},
  \bibinfo{author}{\bibfnamefont{K.}~\bibnamefont{Kamide}},
  \bibinfo{author}{\bibfnamefont{R.}~\bibnamefont{Nii}},
  \bibinfo{author}{\bibfnamefont{T.}~\bibnamefont{Ogawa}}, \bibnamefont{and}
  \bibinfo{author}{\bibfnamefont{Y.}~\bibnamefont{Yamamoto}},
  \bibinfo{journal}{Phys. Rev. Lett.} \textbf{\bibinfo{volume}{111}},
  \bibinfo{pages}{026404} (\bibinfo{year}{2013}),
  \urlprefix\url{https://link.aps.org/doi/10.1103/PhysRevLett.111.026404}.

\bibitem[{\citenamefont{Pertsova and Balatsky}(2020)}]{pertsova2020}
\bibinfo{author}{\bibfnamefont{A.}~\bibnamefont{Pertsova}} \bibnamefont{and}
  \bibinfo{author}{\bibfnamefont{A.~V.} \bibnamefont{Balatsky}},
  \bibinfo{journal}{Annalen der Physik} \textbf{\bibinfo{volume}{532}},
  \bibinfo{pages}{1900549} (\bibinfo{year}{2020}).

\bibitem[{\citenamefont{Christiansen et~al.}(2019)\citenamefont{Christiansen,
  Selig, Malic, Ernstorfer, and Knorr}}]{PhysRevB.100.205401}
\bibinfo{author}{\bibfnamefont{D.}~\bibnamefont{Christiansen}},
  \bibinfo{author}{\bibfnamefont{M.}~\bibnamefont{Selig}},
  \bibinfo{author}{\bibfnamefont{E.}~\bibnamefont{Malic}},
  \bibinfo{author}{\bibfnamefont{R.}~\bibnamefont{Ernstorfer}},
  \bibnamefont{and} \bibinfo{author}{\bibfnamefont{A.}~\bibnamefont{Knorr}},
  \bibinfo{journal}{Phys. Rev. B} \textbf{\bibinfo{volume}{100}},
  \bibinfo{pages}{205401} (\bibinfo{year}{2019}),
  \urlprefix\url{https://link.aps.org/doi/10.1103/PhysRevB.100.205401}.

\bibitem[{\citenamefont{Perfetto
  et~al.}(2020{\natexlab{a}})\citenamefont{Perfetto, Bianchi, and
  Stefanucci}}]{PBS.2020}
\bibinfo{author}{\bibfnamefont{E.}~\bibnamefont{Perfetto}},
  \bibinfo{author}{\bibfnamefont{S.}~\bibnamefont{Bianchi}}, \bibnamefont{and}
  \bibinfo{author}{\bibfnamefont{G.}~\bibnamefont{Stefanucci}},
  \bibinfo{journal}{Phys. Rev. B} \textbf{\bibinfo{volume}{101}},
  \bibinfo{pages}{041201} (\bibinfo{year}{2020}{\natexlab{a}}),
  \urlprefix\url{https://link.aps.org/doi/10.1103/PhysRevB.101.041201}.

\bibitem[{\citenamefont{Mad{\'e}o et~al.}(2020)\citenamefont{Mad{\'e}o, Man,
  Sahoo, Campbell, Pareek, Wong, Mahboob, Chan, Karmakar, Mariserla
  et~al.}}]{madeo2020}
\bibinfo{author}{\bibfnamefont{J.}~\bibnamefont{Mad{\'e}o}},
  \bibinfo{author}{\bibfnamefont{M.~K.} \bibnamefont{Man}},
  \bibinfo{author}{\bibfnamefont{C.}~\bibnamefont{Sahoo}},
  \bibinfo{author}{\bibfnamefont{M.}~\bibnamefont{Campbell}},
  \bibinfo{author}{\bibfnamefont{V.}~\bibnamefont{Pareek}},
  \bibinfo{author}{\bibfnamefont{E.~L.} \bibnamefont{Wong}},
  \bibinfo{author}{\bibfnamefont{A.~A.} \bibnamefont{Mahboob}},
  \bibinfo{author}{\bibfnamefont{N.~S.} \bibnamefont{Chan}},
  \bibinfo{author}{\bibfnamefont{A.}~\bibnamefont{Karmakar}},
  \bibinfo{author}{\bibfnamefont{B.~M.~K.} \bibnamefont{Mariserla}},
  \bibnamefont{et~al.}, \bibinfo{journal}{arXiv preprint arXiv:2005.00241}
  (\bibinfo{year}{2020}).

\bibitem[{\citenamefont{Lee et~al.}(2020)\citenamefont{Lee, Lin, Lu, Chueh,
  Liu, Li, Chang, Kaindl, and Shih}}]{lee2020}
\bibinfo{author}{\bibfnamefont{W.}~\bibnamefont{Lee}},
  \bibinfo{author}{\bibfnamefont{Y.}~\bibnamefont{Lin}},
  \bibinfo{author}{\bibfnamefont{L.-S.} \bibnamefont{Lu}},
  \bibinfo{author}{\bibfnamefont{W.-C.} \bibnamefont{Chueh}},
  \bibinfo{author}{\bibfnamefont{M.}~\bibnamefont{Liu}},
  \bibinfo{author}{\bibfnamefont{X.}~\bibnamefont{Li}},
  \bibinfo{author}{\bibfnamefont{W.-H.} \bibnamefont{Chang}},
  \bibinfo{author}{\bibfnamefont{R.~A.} \bibnamefont{Kaindl}},
  \bibnamefont{and} \bibinfo{author}{\bibfnamefont{C.-K.} \bibnamefont{Shih}},
  \bibinfo{journal}{arXiv preprint arXiv:2008.06103}  (\bibinfo{year}{2020}).

\bibitem[{\citenamefont{Koch et~al.}(2006)\citenamefont{Koch, Kira, Khitrova,
  and Gibbs}}]{Koch2006}
\bibinfo{author}{\bibfnamefont{S.~W.} \bibnamefont{Koch}},
  \bibinfo{author}{\bibfnamefont{M.}~\bibnamefont{Kira}},
  \bibinfo{author}{\bibfnamefont{G.}~\bibnamefont{Khitrova}}, \bibnamefont{and}
  \bibinfo{author}{\bibfnamefont{H.~M.} \bibnamefont{Gibbs}},
  \bibinfo{journal}{Nature Materials} \textbf{\bibinfo{volume}{5}},
  \bibinfo{pages}{523} (\bibinfo{year}{2006}),
  \urlprefix\url{https://doi.org/10.1038/nmat1658}.

\bibitem[{\citenamefont{Madrid et~al.}(2009)\citenamefont{Madrid, Hyeon-Deuk,
  Habenicht, and Prezhdo}}]{madrid2009}
\bibinfo{author}{\bibfnamefont{A.~B.} \bibnamefont{Madrid}},
  \bibinfo{author}{\bibfnamefont{K.}~\bibnamefont{Hyeon-Deuk}},
  \bibinfo{author}{\bibfnamefont{B.~F.} \bibnamefont{Habenicht}},
  \bibnamefont{and} \bibinfo{author}{\bibfnamefont{O.~V.}
  \bibnamefont{Prezhdo}}, \bibinfo{journal}{ACS nano}
  \textbf{\bibinfo{volume}{3}}, \bibinfo{pages}{2487} (\bibinfo{year}{2009}).

\bibitem[{\citenamefont{Nie et~al.}(2014)\citenamefont{Nie, Long, Sun, Huang,
  Zhang, Xiong, Hewak, Shen, Prezhdo, and Loh}}]{nie2014}
\bibinfo{author}{\bibfnamefont{Z.}~\bibnamefont{Nie}},
  \bibinfo{author}{\bibfnamefont{R.}~\bibnamefont{Long}},
  \bibinfo{author}{\bibfnamefont{L.}~\bibnamefont{Sun}},
  \bibinfo{author}{\bibfnamefont{C.-C.} \bibnamefont{Huang}},
  \bibinfo{author}{\bibfnamefont{J.}~\bibnamefont{Zhang}},
  \bibinfo{author}{\bibfnamefont{Q.}~\bibnamefont{Xiong}},
  \bibinfo{author}{\bibfnamefont{D.~W.} \bibnamefont{Hewak}},
  \bibinfo{author}{\bibfnamefont{Z.}~\bibnamefont{Shen}},
  \bibinfo{author}{\bibfnamefont{O.~V.} \bibnamefont{Prezhdo}},
  \bibnamefont{and} \bibinfo{author}{\bibfnamefont{Z.-H.} \bibnamefont{Loh}},
  \bibinfo{journal}{ACS nano} \textbf{\bibinfo{volume}{8}},
  \bibinfo{pages}{10931} (\bibinfo{year}{2014}).

\bibitem[{\citenamefont{Selig et~al.}(2016)\citenamefont{Selig, Bergh{\"a}user,
  Raja, Nagler, Sch{\"u}ller, Heinz, Korn, Chernikov, Malic, and
  Knorr}}]{Selig2016}
\bibinfo{author}{\bibfnamefont{M.}~\bibnamefont{Selig}},
  \bibinfo{author}{\bibfnamefont{G.}~\bibnamefont{Bergh{\"a}user}},
  \bibinfo{author}{\bibfnamefont{A.}~\bibnamefont{Raja}},
  \bibinfo{author}{\bibfnamefont{P.}~\bibnamefont{Nagler}},
  \bibinfo{author}{\bibfnamefont{C.}~\bibnamefont{Sch{\"u}ller}},
  \bibinfo{author}{\bibfnamefont{T.~F.} \bibnamefont{Heinz}},
  \bibinfo{author}{\bibfnamefont{T.}~\bibnamefont{Korn}},
  \bibinfo{author}{\bibfnamefont{A.}~\bibnamefont{Chernikov}},
  \bibinfo{author}{\bibfnamefont{E.}~\bibnamefont{Malic}}, \bibnamefont{and}
  \bibinfo{author}{\bibfnamefont{A.}~\bibnamefont{Knorr}},
  \bibinfo{journal}{Nature Communications} \textbf{\bibinfo{volume}{7}},
  \bibinfo{pages}{13279} (\bibinfo{year}{2016}),
  \urlprefix\url{https://doi.org/10.1038/ncomms13279}.

\bibitem[{\citenamefont{Sangalli et~al.}(2018)\citenamefont{Sangalli, Perfetto,
  Stefanucci, and Marini}}]{sangalli2018ab}
\bibinfo{author}{\bibfnamefont{D.}~\bibnamefont{Sangalli}},
  \bibinfo{author}{\bibfnamefont{E.}~\bibnamefont{Perfetto}},
  \bibinfo{author}{\bibfnamefont{G.}~\bibnamefont{Stefanucci}},
  \bibnamefont{and} \bibinfo{author}{\bibfnamefont{A.}~\bibnamefont{Marini}},
  \bibinfo{journal}{The European Physical Journal B}
  \textbf{\bibinfo{volume}{91}}, \bibinfo{pages}{171} (\bibinfo{year}{2018}).

\bibitem[{\citenamefont{Chernikov
  et~al.}(2015{\natexlab{a}})\citenamefont{Chernikov, van~der Zande, Hill,
  Rigosi, Velauthapillai, Hone, and Heinz}}]{Chernikov2015}
\bibinfo{author}{\bibfnamefont{A.}~\bibnamefont{Chernikov}},
  \bibinfo{author}{\bibfnamefont{A.~M.} \bibnamefont{van~der Zande}},
  \bibinfo{author}{\bibfnamefont{H.~M.} \bibnamefont{Hill}},
  \bibinfo{author}{\bibfnamefont{A.~F.} \bibnamefont{Rigosi}},
  \bibinfo{author}{\bibfnamefont{A.}~\bibnamefont{Velauthapillai}},
  \bibinfo{author}{\bibfnamefont{J.}~\bibnamefont{Hone}}, \bibnamefont{and}
  \bibinfo{author}{\bibfnamefont{T.~F.} \bibnamefont{Heinz}},
  \bibinfo{journal}{Phys. Rev. Lett.} \textbf{\bibinfo{volume}{115}},
  \bibinfo{pages}{126802} (\bibinfo{year}{2015}{\natexlab{a}}),
  \urlprefix\url{https://link.aps.org/doi/10.1103/PhysRevLett.115.126802}.

\bibitem[{\citenamefont{Chernikov
  et~al.}(2015{\natexlab{b}})\citenamefont{Chernikov, Ruppert, Hill, Rigosi,
  and Heinz}}]{Chernikov20152}
\bibinfo{author}{\bibfnamefont{A.}~\bibnamefont{Chernikov}},
  \bibinfo{author}{\bibfnamefont{C.}~\bibnamefont{Ruppert}},
  \bibinfo{author}{\bibfnamefont{H.~M.} \bibnamefont{Hill}},
  \bibinfo{author}{\bibfnamefont{A.~F.} \bibnamefont{Rigosi}},
  \bibnamefont{and} \bibinfo{author}{\bibfnamefont{T.~F.} \bibnamefont{Heinz}},
  \bibinfo{journal}{Nature Photonics} \textbf{\bibinfo{volume}{9}},
  \bibinfo{pages}{466} (\bibinfo{year}{2015}{\natexlab{b}}),
  \urlprefix\url{https://doi.org/10.1038/nphoton.2015.104}.

\bibitem[{\citenamefont{Cunningham et~al.}(2017)\citenamefont{Cunningham,
  Hanbicki, McCreary, and Jonker}}]{Cunningham2017}
\bibinfo{author}{\bibfnamefont{P.~D.} \bibnamefont{Cunningham}},
  \bibinfo{author}{\bibfnamefont{A.~T.} \bibnamefont{Hanbicki}},
  \bibinfo{author}{\bibfnamefont{K.~M.} \bibnamefont{McCreary}},
  \bibnamefont{and} \bibinfo{author}{\bibfnamefont{B.~T.}
  \bibnamefont{Jonker}}, \bibinfo{journal}{ACS Nano}
  \textbf{\bibinfo{volume}{11}}, \bibinfo{pages}{12601} (\bibinfo{year}{2017}),
  \urlprefix\url{https://doi.org/10.1021/acsnano.7b06885}.

\bibitem[{\citenamefont{Yao et~al.}(2017)\citenamefont{Yao, Yan, Kahn, Suslu,
  Liang, Barnard, Tongay, Zettl, Borys, and Schuck}}]{Yao2017}
\bibinfo{author}{\bibfnamefont{K.}~\bibnamefont{Yao}},
  \bibinfo{author}{\bibfnamefont{A.}~\bibnamefont{Yan}},
  \bibinfo{author}{\bibfnamefont{S.}~\bibnamefont{Kahn}},
  \bibinfo{author}{\bibfnamefont{A.}~\bibnamefont{Suslu}},
  \bibinfo{author}{\bibfnamefont{Y.}~\bibnamefont{Liang}},
  \bibinfo{author}{\bibfnamefont{E.~S.} \bibnamefont{Barnard}},
  \bibinfo{author}{\bibfnamefont{S.}~\bibnamefont{Tongay}},
  \bibinfo{author}{\bibfnamefont{A.}~\bibnamefont{Zettl}},
  \bibinfo{author}{\bibfnamefont{N.~J.} \bibnamefont{Borys}}, \bibnamefont{and}
  \bibinfo{author}{\bibfnamefont{P.~J.} \bibnamefont{Schuck}},
  \bibinfo{journal}{Phys. Rev. Lett.} \textbf{\bibinfo{volume}{119}},
  \bibinfo{pages}{087401} (\bibinfo{year}{2017}),
  \urlprefix\url{https://link.aps.org/doi/10.1103/PhysRevLett.119.087401}.

\bibitem[{\citenamefont{Wang et~al.}(2019)\citenamefont{Wang, Ardelean, Bai,
  Steinhoff, Florian, Jahnke, Xu, Kira, Hone, and Zhu}}]{WangArdelean2019}
\bibinfo{author}{\bibfnamefont{J.}~\bibnamefont{Wang}},
  \bibinfo{author}{\bibfnamefont{J.}~\bibnamefont{Ardelean}},
  \bibinfo{author}{\bibfnamefont{Y.}~\bibnamefont{Bai}},
  \bibinfo{author}{\bibfnamefont{A.}~\bibnamefont{Steinhoff}},
  \bibinfo{author}{\bibfnamefont{M.}~\bibnamefont{Florian}},
  \bibinfo{author}{\bibfnamefont{F.}~\bibnamefont{Jahnke}},
  \bibinfo{author}{\bibfnamefont{X.}~\bibnamefont{Xu}},
  \bibinfo{author}{\bibfnamefont{M.}~\bibnamefont{Kira}},
  \bibinfo{author}{\bibfnamefont{J.}~\bibnamefont{Hone}}, \bibnamefont{and}
  \bibinfo{author}{\bibfnamefont{X.-Y.} \bibnamefont{Zhu}},
  \bibinfo{journal}{Science Advances} \textbf{\bibinfo{volume}{5}}
  (\bibinfo{year}{2019}),
  \urlprefix\url{https://advances.sciencemag.org/content/5/9/eaax0145}.

\bibitem[{\citenamefont{Dendzik et~al.}(2020)\citenamefont{Dendzik, Xian,
  Perfetto, Sangalli, Kutnyakhov, Dong, Beaulieu, Pincelli, Pressacco, Curcio
  et~al.}}]{Dendzik2020prl}
\bibinfo{author}{\bibfnamefont{M.}~\bibnamefont{Dendzik}},
  \bibinfo{author}{\bibfnamefont{R.~P.} \bibnamefont{Xian}},
  \bibinfo{author}{\bibfnamefont{E.}~\bibnamefont{Perfetto}},
  \bibinfo{author}{\bibfnamefont{D.}~\bibnamefont{Sangalli}},
  \bibinfo{author}{\bibfnamefont{D.}~\bibnamefont{Kutnyakhov}},
  \bibinfo{author}{\bibfnamefont{S.}~\bibnamefont{Dong}},
  \bibinfo{author}{\bibfnamefont{S.}~\bibnamefont{Beaulieu}},
  \bibinfo{author}{\bibfnamefont{T.}~\bibnamefont{Pincelli}},
  \bibinfo{author}{\bibfnamefont{F.}~\bibnamefont{Pressacco}},
  \bibinfo{author}{\bibfnamefont{D.}~\bibnamefont{Curcio}},
  \bibnamefont{et~al.}, \bibinfo{journal}{Phys. Rev. Lett.}
  \textbf{\bibinfo{volume}{125}}, \bibinfo{pages}{096401}
  (\bibinfo{year}{2020}),
  \urlprefix\url{https://link.aps.org/doi/10.1103/PhysRevLett.125.096401}.

\bibitem[{\citenamefont{Perfetto et~al.}(2016)\citenamefont{Perfetto, Sangalli,
  Marini, and Stefanucci}}]{PSMS.2016}
\bibinfo{author}{\bibfnamefont{E.}~\bibnamefont{Perfetto}},
  \bibinfo{author}{\bibfnamefont{D.}~\bibnamefont{Sangalli}},
  \bibinfo{author}{\bibfnamefont{A.}~\bibnamefont{Marini}}, \bibnamefont{and}
  \bibinfo{author}{\bibfnamefont{G.}~\bibnamefont{Stefanucci}},
  \bibinfo{journal}{Phys. Rev. B} \textbf{\bibinfo{volume}{94}},
  \bibinfo{pages}{245303} (\bibinfo{year}{2016}),
  \urlprefix\url{https://link.aps.org/doi/10.1103/PhysRevB.94.245303}.

\bibitem[{\citenamefont{Steinhoff et~al.}(2017)\citenamefont{Steinhoff,
  Florian, R{\"o}sner, Sch{\"o}nhoff, Wehling, and Jahnke}}]{Steinhoff2017}
\bibinfo{author}{\bibfnamefont{A.}~\bibnamefont{Steinhoff}},
  \bibinfo{author}{\bibfnamefont{M.}~\bibnamefont{Florian}},
  \bibinfo{author}{\bibfnamefont{M.}~\bibnamefont{R{\"o}sner}},
  \bibinfo{author}{\bibfnamefont{G.}~\bibnamefont{Sch{\"o}nhoff}},
  \bibinfo{author}{\bibfnamefont{T.~O.} \bibnamefont{Wehling}},
  \bibnamefont{and} \bibinfo{author}{\bibfnamefont{F.}~\bibnamefont{Jahnke}},
  \bibinfo{journal}{Nature Communications} \textbf{\bibinfo{volume}{8}},
  \bibinfo{pages}{1166} (\bibinfo{year}{2017}),
  \urlprefix\url{https://doi.org/10.1038/s41467-017-01298-6}.

\bibitem[{\citenamefont{Rustagi and Kemper}(2018)}]{RustagiKemper2018}
\bibinfo{author}{\bibfnamefont{A.}~\bibnamefont{Rustagi}} \bibnamefont{and}
  \bibinfo{author}{\bibfnamefont{A.~F.} \bibnamefont{Kemper}},
  \bibinfo{journal}{Phys. Rev. B} \textbf{\bibinfo{volume}{97}},
  \bibinfo{pages}{235310} (\bibinfo{year}{2018}),
  \urlprefix\url{https://link.aps.org/doi/10.1103/PhysRevB.97.235310}.

\bibitem[{\citenamefont{Jiang}(2012)}]{Jiang2012}
\bibinfo{author}{\bibfnamefont{H.}~\bibnamefont{Jiang}}, \bibinfo{journal}{The
  Journal of Physical Chemistry C} \textbf{\bibinfo{volume}{116}},
  \bibinfo{pages}{7664} (\bibinfo{year}{2012}),
  \eprint{https://doi.org/10.1021/jp300079d},
  \urlprefix\url{https://doi.org/10.1021/jp300079d}.

\bibitem[{\citenamefont{Beal and Liang}(1976)}]{beal1976}
\bibinfo{author}{\bibfnamefont{A.}~\bibnamefont{Beal}} \bibnamefont{and}
  \bibinfo{author}{\bibfnamefont{W.}~\bibnamefont{Liang}},
  \bibinfo{journal}{Journal of Physics C: Solid State Physics}
  \textbf{\bibinfo{volume}{9}}, \bibinfo{pages}{2459} (\bibinfo{year}{1976}).

\bibitem[{\citenamefont{Finteis et~al.}(1997)\citenamefont{Finteis,
  Hengsberger, Straub, Fauth, Claessen, Auer, Steiner, H\"ufner, Blaha, V\"ogt
  et~al.}}]{finteis1997}
\bibinfo{author}{\bibfnamefont{T.}~\bibnamefont{Finteis}},
  \bibinfo{author}{\bibfnamefont{M.}~\bibnamefont{Hengsberger}},
  \bibinfo{author}{\bibfnamefont{T.}~\bibnamefont{Straub}},
  \bibinfo{author}{\bibfnamefont{K.}~\bibnamefont{Fauth}},
  \bibinfo{author}{\bibfnamefont{R.}~\bibnamefont{Claessen}},
  \bibinfo{author}{\bibfnamefont{P.}~\bibnamefont{Auer}},
  \bibinfo{author}{\bibfnamefont{P.}~\bibnamefont{Steiner}},
  \bibinfo{author}{\bibfnamefont{S.}~\bibnamefont{H\"ufner}},
  \bibinfo{author}{\bibfnamefont{P.}~\bibnamefont{Blaha}},
  \bibinfo{author}{\bibfnamefont{M.}~\bibnamefont{V\"ogt}},
  \bibnamefont{et~al.}, \bibinfo{journal}{Phys. Rev. B}
  \textbf{\bibinfo{volume}{55}}, \bibinfo{pages}{10400} (\bibinfo{year}{1997}),
  \urlprefix\url{https://link.aps.org/doi/10.1103/PhysRevB.55.10400}.

\bibitem[{\citenamefont{Arora et~al.}(2015)\citenamefont{Arora, Koperski,
  Nogajewski, Marcus, Faugeras, and Potemski}}]{arora2015}
\bibinfo{author}{\bibfnamefont{A.}~\bibnamefont{Arora}},
  \bibinfo{author}{\bibfnamefont{M.}~\bibnamefont{Koperski}},
  \bibinfo{author}{\bibfnamefont{K.}~\bibnamefont{Nogajewski}},
  \bibinfo{author}{\bibfnamefont{J.}~\bibnamefont{Marcus}},
  \bibinfo{author}{\bibfnamefont{C.}~\bibnamefont{Faugeras}}, \bibnamefont{and}
  \bibinfo{author}{\bibfnamefont{M.}~\bibnamefont{Potemski}},
  \bibinfo{journal}{Nanoscale} \textbf{\bibinfo{volume}{7}},
  \bibinfo{pages}{10421} (\bibinfo{year}{2015}).

\bibitem[{\citenamefont{Riley et~al.}(2015)\citenamefont{Riley, Meevasana,
  Bawden, Asakawa, Takayama, Eknapakul, Kim, Hoesch, Mo, Takagi
  et~al.}}]{riley2015}
\bibinfo{author}{\bibfnamefont{J.~M.} \bibnamefont{Riley}},
  \bibinfo{author}{\bibfnamefont{W.}~\bibnamefont{Meevasana}},
  \bibinfo{author}{\bibfnamefont{L.}~\bibnamefont{Bawden}},
  \bibinfo{author}{\bibfnamefont{M.}~\bibnamefont{Asakawa}},
  \bibinfo{author}{\bibfnamefont{T.}~\bibnamefont{Takayama}},
  \bibinfo{author}{\bibfnamefont{T.}~\bibnamefont{Eknapakul}},
  \bibinfo{author}{\bibfnamefont{T.}~\bibnamefont{Kim}},
  \bibinfo{author}{\bibfnamefont{M.}~\bibnamefont{Hoesch}},
  \bibinfo{author}{\bibfnamefont{S.-K.} \bibnamefont{Mo}},
  \bibinfo{author}{\bibfnamefont{H.}~\bibnamefont{Takagi}},
  \bibnamefont{et~al.}, \bibinfo{journal}{Nature nanotechnology}
  \textbf{\bibinfo{volume}{10}}, \bibinfo{pages}{1043} (\bibinfo{year}{2015}).

\bibitem[{\citenamefont{Kim et~al.}(2016)\citenamefont{Kim, Rhim, Kim, Kim, and
  Park}}]{kim2016}
\bibinfo{author}{\bibfnamefont{B.~S.} \bibnamefont{Kim}},
  \bibinfo{author}{\bibfnamefont{J.-W.} \bibnamefont{Rhim}},
  \bibinfo{author}{\bibfnamefont{B.}~\bibnamefont{Kim}},
  \bibinfo{author}{\bibfnamefont{C.}~\bibnamefont{Kim}}, \bibnamefont{and}
  \bibinfo{author}{\bibfnamefont{S.~R.} \bibnamefont{Park}},
  \bibinfo{journal}{Scientific reports} \textbf{\bibinfo{volume}{6}},
  \bibinfo{pages}{36389} (\bibinfo{year}{2016}).

\bibitem[{\citenamefont{Frindt}(1963)}]{frindt1963}
\bibinfo{author}{\bibfnamefont{R.}~\bibnamefont{Frindt}},
  \bibinfo{journal}{Journal of Physics and Chemistry of Solids}
  \textbf{\bibinfo{volume}{24}}, \bibinfo{pages}{1107} (\bibinfo{year}{1963}).

\bibitem[{\citenamefont{Wang et~al.}(2018)\citenamefont{Wang, Chernikov,
  Glazov, Heinz, Marie, Amand, and Urbaszek}}]{WangChernikov2018}
\bibinfo{author}{\bibfnamefont{G.}~\bibnamefont{Wang}},
  \bibinfo{author}{\bibfnamefont{A.}~\bibnamefont{Chernikov}},
  \bibinfo{author}{\bibfnamefont{M.~M.} \bibnamefont{Glazov}},
  \bibinfo{author}{\bibfnamefont{T.~F.} \bibnamefont{Heinz}},
  \bibinfo{author}{\bibfnamefont{X.}~\bibnamefont{Marie}},
  \bibinfo{author}{\bibfnamefont{T.}~\bibnamefont{Amand}}, \bibnamefont{and}
  \bibinfo{author}{\bibfnamefont{B.}~\bibnamefont{Urbaszek}},
  \bibinfo{journal}{Rev. Mod. Phys.} \textbf{\bibinfo{volume}{90}},
  \bibinfo{pages}{021001} (\bibinfo{year}{2018}),
  \urlprefix\url{https://link.aps.org/doi/10.1103/RevModPhys.90.021001}.

\bibitem[{\citenamefont{Bertoni et~al.}(2016)\citenamefont{Bertoni, Nicholson,
  Waldecker, H\"ubener, Monney, De~Giovannini, Puppin, Hoesch, Springate,
  Chapman et~al.}}]{Bertoni_PhysRevLett.117.277201}
\bibinfo{author}{\bibfnamefont{R.}~\bibnamefont{Bertoni}},
  \bibinfo{author}{\bibfnamefont{C.~W.} \bibnamefont{Nicholson}},
  \bibinfo{author}{\bibfnamefont{L.}~\bibnamefont{Waldecker}},
  \bibinfo{author}{\bibfnamefont{H.}~\bibnamefont{H\"ubener}},
  \bibinfo{author}{\bibfnamefont{C.}~\bibnamefont{Monney}},
  \bibinfo{author}{\bibfnamefont{U.}~\bibnamefont{De~Giovannini}},
  \bibinfo{author}{\bibfnamefont{M.}~\bibnamefont{Puppin}},
  \bibinfo{author}{\bibfnamefont{M.}~\bibnamefont{Hoesch}},
  \bibinfo{author}{\bibfnamefont{E.}~\bibnamefont{Springate}},
  \bibinfo{author}{\bibfnamefont{R.~T.} \bibnamefont{Chapman}},
  \bibnamefont{et~al.}, \bibinfo{journal}{Phys. Rev. Lett.}
  \textbf{\bibinfo{volume}{117}}, \bibinfo{pages}{277201}
  (\bibinfo{year}{2016}),
  \urlprefix\url{https://link.aps.org/doi/10.1103/PhysRevLett.117.277201}.

\bibitem[{\citenamefont{Steinhoff et~al.}(2014)\citenamefont{Steinhoff,
  Rösner, Jahnke, Wehling, and Gies}}]{Steinhoff2014}
\bibinfo{author}{\bibfnamefont{A.}~\bibnamefont{Steinhoff}},
  \bibinfo{author}{\bibfnamefont{M.}~\bibnamefont{Rösner}},
  \bibinfo{author}{\bibfnamefont{F.}~\bibnamefont{Jahnke}},
  \bibinfo{author}{\bibfnamefont{T.~O.} \bibnamefont{Wehling}},
  \bibnamefont{and} \bibinfo{author}{\bibfnamefont{C.}~\bibnamefont{Gies}},
  \bibinfo{journal}{Nano Letters} \textbf{\bibinfo{volume}{14}},
  \bibinfo{pages}{3743} (\bibinfo{year}{2014}),
  \urlprefix\url{https://doi.org/10.1021/nl500595u}.

\bibitem[{\citenamefont{Liang and Yang}(2015)}]{Liang2015}
\bibinfo{author}{\bibfnamefont{Y.}~\bibnamefont{Liang}} \bibnamefont{and}
  \bibinfo{author}{\bibfnamefont{L.}~\bibnamefont{Yang}},
  \bibinfo{journal}{Phys. Rev. Lett.} \textbf{\bibinfo{volume}{114}},
  \bibinfo{pages}{063001} (\bibinfo{year}{2015}),
  \urlprefix\url{https://link.aps.org/doi/10.1103/PhysRevLett.114.063001}.

\bibitem[{\citenamefont{Meckbach et~al.}(2018)\citenamefont{Meckbach,
  Stroucken, and Koch}}]{Meckbach2018}
\bibinfo{author}{\bibfnamefont{L.}~\bibnamefont{Meckbach}},
  \bibinfo{author}{\bibfnamefont{T.}~\bibnamefont{Stroucken}},
  \bibnamefont{and} \bibinfo{author}{\bibfnamefont{S.~W.} \bibnamefont{Koch}},
  \bibinfo{journal}{Applied Physics Letters} \textbf{\bibinfo{volume}{112}},
  \bibinfo{pages}{061104} (\bibinfo{year}{2018}),
  \urlprefix\url{https://doi.org/10.1063/1.5017069}.

\bibitem[{\citenamefont{Puppin}(2018)}]{Puppin-PhD}
\bibinfo{author}{\bibfnamefont{M.}~\bibnamefont{Puppin}}, \bibinfo{journal}{Phd
  thesis}  (\bibinfo{year}{2018}),
  \urlprefix\url{http://dx.doi.org/10.17169/refubium-804}.

\bibitem[{\citenamefont{Latini et~al.}(2015)\citenamefont{Latini, Olsen, and
  Thygesen}}]{PhysRevB.92.245123}
\bibinfo{author}{\bibfnamefont{S.}~\bibnamefont{Latini}},
  \bibinfo{author}{\bibfnamefont{T.}~\bibnamefont{Olsen}}, \bibnamefont{and}
  \bibinfo{author}{\bibfnamefont{K.~S.} \bibnamefont{Thygesen}},
  \bibinfo{journal}{Phys. Rev. B} \textbf{\bibinfo{volume}{92}},
  \bibinfo{pages}{245123} (\bibinfo{year}{2015}),
  \urlprefix\url{https://link.aps.org/doi/10.1103/PhysRevB.92.245123}.

\bibitem[{\citenamefont{Wallauer et~al.}(2016)\citenamefont{Wallauer, Reimann,
  Armbrust, G{\"u}dde, and H{\"o}fer}}]{wallauer2016}
\bibinfo{author}{\bibfnamefont{R.}~\bibnamefont{Wallauer}},
  \bibinfo{author}{\bibfnamefont{J.}~\bibnamefont{Reimann}},
  \bibinfo{author}{\bibfnamefont{N.}~\bibnamefont{Armbrust}},
  \bibinfo{author}{\bibfnamefont{J.}~\bibnamefont{G{\"u}dde}},
  \bibnamefont{and}
  \bibinfo{author}{\bibfnamefont{U.}~\bibnamefont{H{\"o}fer}},
  \bibinfo{journal}{Applied Physics Letters} \textbf{\bibinfo{volume}{109}},
  \bibinfo{pages}{162102} (\bibinfo{year}{2016}).

\bibitem[{\citenamefont{Waldecker et~al.}(2017)\citenamefont{Waldecker,
  Bertoni, H{\"u}bener, Brumme, Vasileiadis, Zahn, Rubio, and
  Ernstorfer}}]{waldecker2017}
\bibinfo{author}{\bibfnamefont{L.}~\bibnamefont{Waldecker}},
  \bibinfo{author}{\bibfnamefont{R.}~\bibnamefont{Bertoni}},
  \bibinfo{author}{\bibfnamefont{H.}~\bibnamefont{H{\"u}bener}},
  \bibinfo{author}{\bibfnamefont{T.}~\bibnamefont{Brumme}},
  \bibinfo{author}{\bibfnamefont{T.}~\bibnamefont{Vasileiadis}},
  \bibinfo{author}{\bibfnamefont{D.}~\bibnamefont{Zahn}},
  \bibinfo{author}{\bibfnamefont{A.}~\bibnamefont{Rubio}}, \bibnamefont{and}
  \bibinfo{author}{\bibfnamefont{R.}~\bibnamefont{Ernstorfer}},
  \bibinfo{journal}{Physical Review Letters} \textbf{\bibinfo{volume}{119}},
  \bibinfo{pages}{036803} (\bibinfo{year}{2017}).

\bibitem[{\citenamefont{Haug et~al.}(1994)\citenamefont{Haug, , and
  Koch}}]{HaugKochbook}
\bibinfo{author}{\bibfnamefont{H.}~\bibnamefont{Haug}}, , \bibnamefont{and}
  \bibinfo{author}{\bibfnamefont{S.~W.} \bibnamefont{Koch}},
  \emph{\bibinfo{title}{Quantum Theory of the Optical and Electronic Properties
  of Semiconductors}} (\bibinfo{publisher}{World Scientific},
  \bibinfo{address}{Singapore}, \bibinfo{year}{1994}).

\bibitem[{\citenamefont{Steinhoff et~al.}(2016)\citenamefont{Steinhoff,
  Florian, Rösner, Lorke, Wehling, Gies, and Jahnke}}]{Steinhoff2016}
\bibinfo{author}{\bibfnamefont{A.}~\bibnamefont{Steinhoff}},
  \bibinfo{author}{\bibfnamefont{M.}~\bibnamefont{Florian}},
  \bibinfo{author}{\bibfnamefont{M.}~\bibnamefont{Rösner}},
  \bibinfo{author}{\bibfnamefont{M.}~\bibnamefont{Lorke}},
  \bibinfo{author}{\bibfnamefont{T.~O.} \bibnamefont{Wehling}},
  \bibinfo{author}{\bibfnamefont{C.}~\bibnamefont{Gies}}, \bibnamefont{and}
  \bibinfo{author}{\bibfnamefont{F.}~\bibnamefont{Jahnke}},
  \bibinfo{journal}{2D Materials} \textbf{\bibinfo{volume}{3}},
  \bibinfo{pages}{031006} (\bibinfo{year}{2016}),
  \urlprefix\url{https://doi.org/10.1088%2F2053-1583%2F3%2F3%2F031006}.

\bibitem[{\citenamefont{Schmidt et~al.}(2016)\citenamefont{Schmidt, Berghauser,
  Schneider, Selig, Tonndorf, Malic, Knorr, Michaelis~de Vasconcellos, and
  Bratschitsch}}]{schmidt2016}
\bibinfo{author}{\bibfnamefont{R.}~\bibnamefont{Schmidt}},
  \bibinfo{author}{\bibfnamefont{G.}~\bibnamefont{Berghauser}},
  \bibinfo{author}{\bibfnamefont{R.}~\bibnamefont{Schneider}},
  \bibinfo{author}{\bibfnamefont{M.}~\bibnamefont{Selig}},
  \bibinfo{author}{\bibfnamefont{P.}~\bibnamefont{Tonndorf}},
  \bibinfo{author}{\bibfnamefont{E.}~\bibnamefont{Malic}},
  \bibinfo{author}{\bibfnamefont{A.}~\bibnamefont{Knorr}},
  \bibinfo{author}{\bibfnamefont{S.}~\bibnamefont{Michaelis~de Vasconcellos}},
  \bibnamefont{and}
  \bibinfo{author}{\bibfnamefont{R.}~\bibnamefont{Bratschitsch}},
  \bibinfo{journal}{Nano letters} \textbf{\bibinfo{volume}{16}},
  \bibinfo{pages}{2945} (\bibinfo{year}{2016}).

\bibitem[{\citenamefont{Molina-S\'anchez
  et~al.}(2017)\citenamefont{Molina-S\'anchez, Sangalli, Wirtz, and
  Marini}}]{MolinaSanchez2017}
\bibinfo{author}{\bibfnamefont{A.}~\bibnamefont{Molina-S\'anchez}},
  \bibinfo{author}{\bibfnamefont{D.}~\bibnamefont{Sangalli}},
  \bibinfo{author}{\bibfnamefont{L.}~\bibnamefont{Wirtz}}, \bibnamefont{and}
  \bibinfo{author}{\bibfnamefont{A.}~\bibnamefont{Marini}},
  \bibinfo{journal}{Nano Letters} \textbf{\bibinfo{volume}{17}},
  \bibinfo{pages}{4549} (\bibinfo{year}{2017}),
  \urlprefix\url{https://doi.org/10.1021/acs.nanolett.7b00175}.

\bibitem[{\citenamefont{Perfetto
  et~al.}(2020{\natexlab{b}})\citenamefont{Perfetto, Marini, and
  Stefanucci}}]{PhysRevB.102.085203}
\bibinfo{author}{\bibfnamefont{E.}~\bibnamefont{Perfetto}},
  \bibinfo{author}{\bibfnamefont{A.}~\bibnamefont{Marini}}, \bibnamefont{and}
  \bibinfo{author}{\bibfnamefont{G.}~\bibnamefont{Stefanucci}},
  \bibinfo{journal}{Phys. Rev. B} \textbf{\bibinfo{volume}{102}},
  \bibinfo{pages}{085203} (\bibinfo{year}{2020}{\natexlab{b}}),
  \urlprefix\url{https://link.aps.org/doi/10.1103/PhysRevB.102.085203}.

\bibitem[{\citenamefont{Giuliani and Vignale}(2005)}]{GiulianiVignale-book}
\bibinfo{author}{\bibfnamefont{G.}~\bibnamefont{Giuliani}} \bibnamefont{and}
  \bibinfo{author}{\bibfnamefont{G.}~\bibnamefont{Vignale}},
  \emph{\bibinfo{title}{Quantum Theory of the Electron Liquid}}
  (\bibinfo{publisher}{Cambridge University Press},
  \bibinfo{address}{Cambridge}, \bibinfo{year}{2005}).

\bibitem[{\citenamefont{Halperin and Rice}(1968)}]{HalperinRice1968}
\bibinfo{author}{\bibfnamefont{B.}~\bibnamefont{Halperin}} \bibnamefont{and}
  \bibinfo{author}{\bibfnamefont{T.}~\bibnamefont{Rice}},
  \bibinfo{journal}{Solid State Physics} \textbf{\bibinfo{volume}{21}},
  \bibinfo{pages}{115 } (\bibinfo{year}{1968}),
  \urlprefix\url{http://www.sciencedirect.com/science/article/pii/S0081194708607407}.

\bibitem[{\citenamefont{Blatt et~al.}(1962)\citenamefont{Blatt, B\"oer, and
  Brandt}}]{Blatt1962}
\bibinfo{author}{\bibfnamefont{J.~M.} \bibnamefont{Blatt}},
  \bibinfo{author}{\bibfnamefont{K.~W.} \bibnamefont{B\"oer}},
  \bibnamefont{and} \bibinfo{author}{\bibfnamefont{W.}~\bibnamefont{Brandt}},
  \bibinfo{journal}{Phys. Rev.} \textbf{\bibinfo{volume}{126}},
  \bibinfo{pages}{1691} (\bibinfo{year}{1962}),
  \urlprefix\url{https://link.aps.org/doi/10.1103/PhysRev.126.1691}.

\bibitem[{\citenamefont{Keldysh and Kopaev}(1965)}]{KeldyshKopaev1965}
\bibinfo{author}{\bibfnamefont{L.~V.} \bibnamefont{Keldysh}} \bibnamefont{and}
  \bibinfo{author}{\bibfnamefont{Y.~U.} \bibnamefont{Kopaev}},
  \bibinfo{journal}{Sov. Phys. Solid State} \textbf{\bibinfo{volume}{6}},
  \bibinfo{pages}{2219} (\bibinfo{year}{1965}).

\bibitem[{\citenamefont{Kozlov and Maksimov}(1965)}]{Kozlov-Maksimov_JETP1965}
\bibinfo{author}{\bibfnamefont{A.~N.} \bibnamefont{Kozlov}} \bibnamefont{and}
  \bibinfo{author}{\bibfnamefont{L.~A.} \bibnamefont{Maksimov}},
  \bibinfo{journal}{JETP} \textbf{\bibinfo{volume}{21}}, \bibinfo{pages}{790}
  (\bibinfo{year}{1965}).

\bibitem[{\citenamefont{J\'erome et~al.}(1967)\citenamefont{J\'erome, Rice, and
  Kohn}}]{JeromeRiceKohn1967}
\bibinfo{author}{\bibfnamefont{D.}~\bibnamefont{J\'erome}},
  \bibinfo{author}{\bibfnamefont{T.~M.} \bibnamefont{Rice}}, \bibnamefont{and}
  \bibinfo{author}{\bibfnamefont{W.}~\bibnamefont{Kohn}},
  \bibinfo{journal}{Phys. Rev.} \textbf{\bibinfo{volume}{158}},
  \bibinfo{pages}{462} (\bibinfo{year}{1967}),
  \urlprefix\url{https://link.aps.org/doi/10.1103/PhysRev.158.462}.

\bibitem[{\citenamefont{{Comte, C.} and {Nozi\`eres,
  P.}}(1982)}]{Comte-Nozieres_1982}
\bibinfo{author}{\bibnamefont{{Comte, C.}}} \bibnamefont{and}
  \bibinfo{author}{\bibnamefont{{Nozi\`eres, P.}}}, \bibinfo{journal}{J. Phys.
  France} \textbf{\bibinfo{volume}{43}}, \bibinfo{pages}{1069}
  (\bibinfo{year}{1982}),
  \urlprefix\url{https://doi.org/10.1051/jphys:019820043070106900}.

\bibitem[{\citenamefont{Perfetto and
  Stefanucci}(2020)}]{PhysRevLett.125.106401}
\bibinfo{author}{\bibfnamefont{E.}~\bibnamefont{Perfetto}} \bibnamefont{and}
  \bibinfo{author}{\bibfnamefont{G.}~\bibnamefont{Stefanucci}},
  \bibinfo{journal}{Phys. Rev. Lett.} \textbf{\bibinfo{volume}{125}},
  \bibinfo{pages}{106401} (\bibinfo{year}{2020}),
  \urlprefix\url{https://link.aps.org/doi/10.1103/PhysRevLett.125.106401}.

\bibitem[{\citenamefont{Perfetto and Stefanucci}(2018)}]{PS-cheers}
\bibinfo{author}{\bibfnamefont{E.}~\bibnamefont{Perfetto}} \bibnamefont{and}
  \bibinfo{author}{\bibfnamefont{G.}~\bibnamefont{Stefanucci}},
  \bibinfo{journal}{Journal of Physics: Condensed Matter}
  \textbf{\bibinfo{volume}{30}}, \bibinfo{pages}{465901}
  (\bibinfo{year}{2018}),
  \urlprefix\url{http://stacks.iop.org/0953-8984/30/i=46/a=465901}.

\bibitem[{\citenamefont{Sangalli et~al.}(2019)\citenamefont{Sangalli, Ferretti,
  Miranda, Attaccalite, Marri, Cannuccia, Melo, Marsili, Paleari, Marrazzo
  et~al.}}]{Sangalli2019}
\bibinfo{author}{\bibfnamefont{D.}~\bibnamefont{Sangalli}},
  \bibinfo{author}{\bibfnamefont{A.}~\bibnamefont{Ferretti}},
  \bibinfo{author}{\bibfnamefont{H.}~\bibnamefont{Miranda}},
  \bibinfo{author}{\bibfnamefont{C.}~\bibnamefont{Attaccalite}},
  \bibinfo{author}{\bibfnamefont{I.}~\bibnamefont{Marri}},
  \bibinfo{author}{\bibfnamefont{E.}~\bibnamefont{Cannuccia}},
  \bibinfo{author}{\bibfnamefont{P.}~\bibnamefont{Melo}},
  \bibinfo{author}{\bibfnamefont{M.}~\bibnamefont{Marsili}},
  \bibinfo{author}{\bibfnamefont{F.}~\bibnamefont{Paleari}},
  \bibinfo{author}{\bibfnamefont{A.}~\bibnamefont{Marrazzo}},
  \bibnamefont{et~al.}, \bibinfo{journal}{Journal of Physics: Condensed Matter}
  \textbf{\bibinfo{volume}{31}}, \bibinfo{pages}{325902}
  (\bibinfo{year}{2019}),
  \urlprefix\url{https://doi.org/10.1088%2F1361-648x%2Fab15d0}.

\bibitem[{\citenamefont{Freericks et~al.}(2009)\citenamefont{Freericks,
  Krishnamurthy, and Pruschke}}]{Freericks_PhysRevLett.102.136401}
\bibinfo{author}{\bibfnamefont{J.~K.} \bibnamefont{Freericks}},
  \bibinfo{author}{\bibfnamefont{H.~R.} \bibnamefont{Krishnamurthy}},
  \bibnamefont{and} \bibinfo{author}{\bibfnamefont{T.}~\bibnamefont{Pruschke}},
  \bibinfo{journal}{Phys. Rev. Lett.} \textbf{\bibinfo{volume}{102}},
  \bibinfo{pages}{136401} (\bibinfo{year}{2009}),
  \urlprefix\url{https://link.aps.org/doi/10.1103/PhysRevLett.102.136401}.

\bibitem[{\citenamefont{Stefanucci and van Leeuwen}(2013)}]{svl-book}
\bibinfo{author}{\bibfnamefont{G.}~\bibnamefont{Stefanucci}} \bibnamefont{and}
  \bibinfo{author}{\bibfnamefont{R.}~\bibnamefont{van Leeuwen}},
  \emph{\bibinfo{title}{Nonequilibrium Many-Body Theory of Quantum Systems: A
  Modern Introduction}} (\bibinfo{publisher}{Cambridge University Press},
  \bibinfo{address}{Cambridge}, \bibinfo{year}{2013}).

\bibitem[{\citenamefont{Perfetto et~al.}(2015)\citenamefont{Perfetto, Uimonen,
  van Leeuwen, and Stefanucci}}]{PUvLS.2015}
\bibinfo{author}{\bibfnamefont{E.}~\bibnamefont{Perfetto}},
  \bibinfo{author}{\bibfnamefont{A.-M.} \bibnamefont{Uimonen}},
  \bibinfo{author}{\bibfnamefont{R.}~\bibnamefont{van Leeuwen}},
  \bibnamefont{and}
  \bibinfo{author}{\bibfnamefont{G.}~\bibnamefont{Stefanucci}},
  \bibinfo{journal}{Phys. Rev. A} \textbf{\bibinfo{volume}{92}},
  \bibinfo{pages}{033419} (\bibinfo{year}{2015}),
  \urlprefix\url{https://link.aps.org/doi/10.1103/PhysRevA.92.033419}.

\bibitem[{\citenamefont{Lipavsk\'y et~al.}(1986)\citenamefont{Lipavsk\'y,
  \ifmmode \check{S}\else \v{S}\fi{}pi\ifmmode~\check{c}\else \v{c}\fi{}ka, and
  Velick\'y}}]{PhysRevB.34.6933}
\bibinfo{author}{\bibfnamefont{P.}~\bibnamefont{Lipavsk\'y}},
  \bibinfo{author}{\bibfnamefont{V.}~\bibnamefont{\ifmmode \check{S}\else
  \v{S}\fi{}pi\ifmmode~\check{c}\else \v{c}\fi{}ka}}, \bibnamefont{and}
  \bibinfo{author}{\bibfnamefont{B.}~\bibnamefont{Velick\'y}},
  \bibinfo{journal}{Phys. Rev. B} \textbf{\bibinfo{volume}{34}},
  \bibinfo{pages}{6933} (\bibinfo{year}{1986}),
  \urlprefix\url{https://link.aps.org/doi/10.1103/PhysRevB.34.6933}.

\bibitem[{\citenamefont{Latini et~al.}(2014)\citenamefont{Latini, Perfetto,
  Uimonen, van Leeuwen, and Stefanucci}}]{LPUvLS.2014}
\bibinfo{author}{\bibfnamefont{S.}~\bibnamefont{Latini}},
  \bibinfo{author}{\bibfnamefont{E.}~\bibnamefont{Perfetto}},
  \bibinfo{author}{\bibfnamefont{A.-M.} \bibnamefont{Uimonen}},
  \bibinfo{author}{\bibfnamefont{R.}~\bibnamefont{van Leeuwen}},
  \bibnamefont{and}
  \bibinfo{author}{\bibfnamefont{G.}~\bibnamefont{Stefanucci}},
  \bibinfo{journal}{Phys. Rev. B} \textbf{\bibinfo{volume}{89}},
  \bibinfo{pages}{075306} (\bibinfo{year}{2014}),
  \urlprefix\url{https://link.aps.org/doi/10.1103/PhysRevB.89.075306}.

\end{thebibliography}

\end{document}